\documentclass[prb,twocolumn,showpacs,preprintnumbers,amsmath,amssymb,floatfix]{revtex4-1}
\usepackage{graphicx}
\usepackage{bm}\let\vec\bm

\newcommand \be{\begin{eqnarray}}
\newcommand \ee{\end{eqnarray}}
\newcommand \ba{\begin{align}}
\newcommand \eea{\end{align}}

\newcommand {\p}[1]{\partial_{#1}}
\newcommand {\pp}[1]{{\partial\over \partial {#1}}}

\newcommand \V{\vec}

\begin{document}
           \csname @twocolumnfalse\endcsname
\title{Nonequilibrium thermodynamics with binary quantum correlations}
\author{K. Morawetz$^{1,2,3}$
}
\affiliation{$^1$M\"unster University of Applied Sciences,
Stegerwaldstrasse 39, 48565 Steinfurt, Germany}
\affiliation{$^2$International Institute of Physics- UFRN,
Campus Universit\'ario Lagoa nova,
CEP: 59078-970 / C.P. 1613, Natal, Brazil
}
\affiliation{$^{3}$ Max-Planck-Institute for the Physics of Complex Systems, 01187 Dresden, Germany
}

\begin{abstract}
The balance equations for thermodynamic quantities are derived from the nonlocal quantum kinetic equation. The nonlocal collisions lead to molecular contributions to the observables and currents. The corresponding correlated parts of the observables are found to be given by the rate to form a molecule multiplied with its lifetime which can be considered as collision duration. Explicit expressions of these molecular contributions are given in terms of the scattering phase shifts. The two-particle form of the entropy is derived extending the Landau quasiparticle picture by two-particle molecular contributions. There is a continuous exchange of correlation and kinetic energies condensing into the rate of correlated variables for energy and momentum. For the entropy, an explicit gain remains and Boltzmann's H-theorem is proved including the molecular parts of the 
entropy.  
\end{abstract}
\pacs{
05.60.Gg. 
05.70.Ln, 
47.70.Nd,
51.10.+y, 
}
\maketitle

\section{Introduction}

Highly non-equilibrium Fermi systems occur in various fields of physics,
e.g. electrons driven by fast lasers, nucleons in nuclear reactions
or atoms in ultra-cold gases. The dynamics of such systems is often too
complex to be treated by exact quantum statistical approaches. This is caused
by the strong interaction.
A feasible microscopic picture is provided by quasi-classical
simulations of single-particle trajectories in self-consistent
force fields and randomly selected binary collisions. Although
these simulations solve in principle a kinetic equation offering the complete
single-particle distribution in phase space, the main results
are hydrodynamical quantities like the particle flow and the
corresponding density profile, because of their clear interpretation.

The relation between the single-particle distribution and the particle
density is trivial as long as binary collisions are so fast that
they can be treated as instantaneous. If the finite duration of
collisions becomes important, a part of particles is hidden in
collision states found here as molecular states and a more sophisticated evaluation is necessary.
In this paper we evaluate the hydrodynamic and thermodynamic
quantities as functionals of the single-particle nonequilibrium distribution
including nonlocal collisions of finite duration. We derive balance
equations for densities of particles, momentum, energy and the entropy.
As will be seen, the finite duration of collisions leads to
molecular contributions in all balance equations.

\paragraph{History of nonlocal collisions}

The very basic idea of the Boltzmann equation from 1872 \cite{B72}, to balance
the drift of particles with dissipation, is used in all mentioned fields allowing for a number of improvements that make it possible to describe phenomena
far beyond the range of the validity of the original Boltzmann equation. In
these improvements the theory of gases differs from the theory of condensed
systems.

In the theory of gases, the focus was on the so called virial corrections that take
into account a finite volume of molecules and an effective pressure caused by their
interaction. The original Boltzmann equation cannot describe virial corrections
because the instant and local approximation of scattering events
implies an ideal gas equation of state. To extend
the validity of the Boltzmann equation to moderately dense gases, Clausius and
Boltzmann included the space nonlocality of binary collisions \cite{CC90}.
For the model of hard spheres, Enskog \cite{E21} has further extended
the nonlocal collision integrals by statistical correlations.
It was later modified to the nowadays used revised Enskog theory \cite{BE73}.
An effort to describe the virial corrections for real particles, in particular when their de Broglie wave lengths are comparable with the potential range,
has resulted in various generalizations of Enskog's equation
\cite{W57,W58,W60,S60,S64,B69,TS70,SS71,RS76,B75,M89,La89,TNL89,NTL89,L90,L90a,H90,H90a,H91,LM90,S90,S91,NTL91,NTL91a,S95,SML95}.
By closer inspection one finds that all tractable quantum theories deal
exclusively with non-local corrections. The statistical correlations
in quantum systems would require an adequate solution of three-particle
collisions from Fadeev equations \cite{BRS96,PKP00,PP96}. A systematic incorporation of the latter one into the kinetic equation, however, is not yet fully
understood, therefore we discuss only binary processes.

In the theory of condensed systems, a historical headway was
the Landau concept of quasiparticles\cite{BP91} with three major
modifications of the Boltzmann equation: the Pauli blocking of scattering channels, the underlying quantum
mechanical dynamics of collisions, and the single-particle-like excitations
(quasiparticles) instead of real particles. Unlike in the theory of
gases, the scattering integrals of the Boltzmann equation remain local in space
and time and the Landau theory does not include a quantum analog to 
(non-local) virial corrections.


Although nuclear matter is a dense Fermi liquid and very much benefits
from the Landau concept, it was felt that nonlocal contributions are missing.
Attempts started from numerical experiments \cite{H81} within
the cascade model and nonlocal corrections of Enskog type \cite{M83}, 
incorporated into the Monte-Carlo codes for the
Boltzmann equation \cite{KDB96} using a method developed within the classical
molecular dynamics \cite{AGA95}.
These implementations of Enskog's corrections to nuclear reactions did
not improve the agreement with experimental data \cite{BGM94}. 
One of the discussed
reasons for this disagreement were the statistical correlations studied
in detail for classical hard-sphere model\cite{E21,CC90,HCB64,Sch91,C62,W63,W63a,W65,KO65,DC67,GF67,BE79}.

It turned out that the original Enskog corrections are not suited for nuclear
matter, because the dominant correction is the finite duration of the nucleon-nucleon 
collision, not its nonlocality \cite{DP96}. Pioneering simulations were thus heading
in the wrong direction, because the hard spheres lead to the excluded
volume and thus to a compressibility lower than the one of an ideal gas, while the
finite duration increases the compressibility because a fraction of particles
is bounded in short-living molecules.

Collisions with nonlocal corrections obtained from realistic nucleon-nucleon 
scattering phase shifts \cite{MLSK99} were first implemented in 
simulations\cite{MSLKKN99} of nuclear reactions.
As a result, the hydrodynamic properties changes and
a hot neck between two reacting nuclei shows a longer
lifetime which increases the production of hot protons and neutrons reducing
the discrepancy between experimental and simulated data \cite{MLNCCT01}.
It was encouraging that the nonlocal corrections have no effect on the run time
of simulations. This is in contrast with quasiparticle contributions, because
the back-flow (within the Landau local approach) is a hard numerical problem.
Within the nonlocal approach, one can view non-dissipative interactions among
particles as zero-angle `collisions'. Replacing rejected Pauli-blocked collisions
by nonlocal zero-angle collisions, modified velocities of quasiparticles and
back-flows are simulated on no numerical cost \cite{LSM97,MLNCCT01}.

The simulated corrections \cite{LSM97,MLNCCT01} mimic the collision delay 
by seeding particles after the local 
collision into positions from which they would have arrived  asymptotically when colliding with the true delay. Quantum studies of gases based 
on Waldmann's equation\cite{W57} generalized by Snider\cite{S60,S64} and 
further developed by Lalo\"e, Mullin, Nacher and Tastevin\cite{La89,TNL89,NTL89,LM90,NTL91,NTL91a}
directly result in instantaneous collisions with nonlocal corrections modeling 
the collision delay. The invisibility of the collision delay in Snider's
approach follows from the absent off-shell motion during collisions. The
Wigner collision delay is the energy derivative of the scattering phase, 
but Snider's approach provides the scattering phase only for the `on-shell' 
energy equal to the sum of final single-particle energies. The same limitation 
applies to the approach of de Haan\cite{H90,H90a,H91} who confirmed the results of 
Lalo\"e, Nacher and Tastevin using Balescu's formal derivation of kinetic 
equations \cite{B75}.


It should be noted that the approach of Lalo\"e, Tastevin and Nacher is
limited to non-degenerate systems, therefore their quasiparticle
features are given by other than exchange processes. 

\paragraph{Collision delay}
Unlike nonlocal corrections in space, the true collision delay is
still a problem for implementations. In spite of the lack of an 
effective simulation scheme, here we want to discuss properties of 
the kinetic equation with collision delay. In particular in dense 
systems the collision delay has to be treated properly because during 
the collision particles contribute to the background as part of the Pauli 
blocking of states.

Let us outline the quantum collision delay and its interpretation.
In the weak-coupling limit the scattering rate of two particles is given
by the matrix element of the interaction potential between the plain waves
of initial and final states. For strong potentials, the wave function
cannot completely penetrate the core but can be enhanced at moderately short
distances. In the strong-coupling case one thus has to take into
account the reconstruction of the wave functions by the interaction
potential. It reflects the finite duration of the collision usually called 
internal dynamics. The build-up of the wave function means that the
particles can be found at a given distance with an increased
probability. Within the ergodic interpretation of the probability it
means that they have to spend a longer time at this distance than with
an uncorrelated motion corresponding to
the concept of the dwell time \cite{HS89}. A recent review \cite{KV13}
shows numerous definitions of the collision delay and discusses their
relevance to the quantum kinetic theory. The gradient expansion supports
the Wigner delay time as energy derivative of the
scattering phase shift.

Within many-body Green's functions, the space and time nonlocal corrections
are treated on equal footage and the internal dynamics of collisions is
described by the two-particle T-matrix which has been introduced to
describe the reconstruction of the wave function in the collision
process. The physical content of the T-matrix is, however, easily wasted
when one derives the kinetic equation. Most of the Green's function studies
result in Landau's kinetic equation, where the scattering integral is
instant and local. This contradiction follows from the second 'well
established' approximation which is the neglect of gradient corrections to the
scattering integral. The headway for a Green's function treatment of
non-instant and nonlocal corrections to the scattering integral was
done by B{\"a}rwinkel \cite{B69} who also discussed the thermodynamical
consequences of these corrections\cite{B69a}. The present approach is
based on non-equilibrium Green's functions \cite{LSM97,SLM96b} known as 
the generalized Kadanoff-Baym (GKB) formalism taking into account consistently all first-order gradient corrections \cite{MLS00}.

\paragraph{Entropy}
The entropy as a measure of complexity, or inversely as the loss of information,
plays a central role in processes like nuclear or cluster reactions, where the
kinetic and correlation energy of projectile and target particles transform into heat.
In nuclear matter, mainly the single-particle entropy \cite{IKV03,P04,ACHMN08,Mou10,SR14} is 
discussed as it is in ultra-cold atoms \cite{BE10}.
The equilibrium entropy has been given in a form of cluster expansion where the
two-particle part is represented by the two-particle correlation function \cite{K42}
which has been calculated numerically for different systems\cite{LH92,NC13}.
The genuine two-particle part of the quantum entropy is still an open question as well as its nonequilibrium expression. 

Equilibrium values are presented in terms of either general Green's functions \cite{VB98,MT15} or in expansion of coupling parameters \cite{BIR01}. General expressions of the entropy in terms of $\Phi$-derivable functionals \cite{CP75} would require a tremendous reduction in order to understand the contribution of correlated parts and single-parts explicitly in an applicable form.
We proceed another way and employ the nonlocal kinetic equation which contains single and two-particle correlated parts to extract the correlated entropy from balance equations. Therefore we are using the infinite ladder summation condensed in the T-matrix since the correlated entropy is a result essential beyond the one-loop approximation \cite{VB98}. The second advantage of our approach is that we present the nonequilibrium expression of the entropy where no extremal principle for $\Phi$-derivable functionals can be given.

Very 
often the entanglement entropy is also investigated if one set of variables is
traced off from the density operator which provides the information
exchanged between the two subsystems \cite{PE09}. The extraction of two
particles out of a many-body state leads to a different entanglement
entropy \cite{SNB09} than the one of the reduced density matrix. Similarly
the calculation of either spatial-dependent or momentum-dependent one-
and two-particle entropies yields different results \cite{LSTA16}.
 The majority of approaches calculate the
classical entropy in various approximations \cite{Puo99,He00}. Here we
will obtain the quantum one- and two-particle entropies explicitly in terms of phase 
shifts of the scattering T-matrix. 

\paragraph{Outline of the paper}

First we give
the nonlocal kinetic equation derived first in \cite{SLM96,LSM97} and present important symmetries of the collision integral in chapter III. Then we derive the thermodynamic quantities as balance equations for density, momentum, energy, and entropy from this kinetic equation  in chapter IV. We show that besides the usual balance equations for quasiparticles, where the integrals over the collision integral vanishes, the nonlocality of the collision process induces explicit molecular contributions. In Chapter V we summarize the forms of balance equations discussing their statistical interpretation and prove Boltzmann's H-theorem. 
The summary is in chapter VI. 

\section{Nonlocal kinetic theory}

\subsection{Nonlocal kinetic equation}

The nonlocal kinetic equation 
reads
 \begin{equation}
{\partial f_1\over\partial t}+{\partial\varepsilon_1\over\partial \V {k}}
{\partial f_1\over\partial \V{r}}-{\partial\varepsilon_1\over\partial \V{r}}
{\partial f_1\over\partial \V{k}}
=I_1^{\mathrm{in}}-I_1^{\mathrm{out}}
\label{vc17}
\end{equation}
with the scattering-in
\ba
&I_1^{\mathrm{in}}=\sum_b\int {d^3 \! p\over(2\pi)^3}{d^3 \! q\over(2\pi)^3}
2\pi\delta\left(\varepsilon_1+\bar \varepsilon_2-\bar \varepsilon_3
-\bar \varepsilon_4-2\Delta_E\right)
\nonumber\\
&\times
\Biggl(1-{1\over 2}{\partial\V \Delta_2\over\partial \V r}
-{\partial\bar \varepsilon_2\over\partial \V r}
{\partial\V \Delta_2\over\partial\omega}\Biggr)_{\omega=\varepsilon_1+\bar \varepsilon_2}
(1-f_1-\bar {f_2})\bar {f_3} \bar {f_4}
\nonumber\\
&\times
\!\left|t_{\rm sc}\!\Bigl(\!\varepsilon_1\!+\!\bar {\varepsilon}_2\!-\!
\Delta_E,\V k\!-\!{\V \Delta_K\over 2},\!\V p-\!{\V \Delta_K\over 2},\V q,\V r\!-\!
{\V \Delta_r},t\!-\!{\Delta_t\over 2}\!\Bigr)\!\right|^2
\label{Iin}
\end{align}
and the scattering-out
\ba
&
I_1^{\mathrm{out}}=
\sum_b\int {d^3 \! p\over(2\pi)^3}{d^3 \! q\over(2\pi)^3}
2\pi\delta\left(\varepsilon_1+\varepsilon_2-\varepsilon_3
-\varepsilon_4+2\Delta_E\right)
\nonumber\\
&\times
\Biggl(1+{1\over 2}{\partial\V \Delta_2\over\partial \V r}
+{\partial\varepsilon_2\over\partial r}
{\partial\V \Delta_2\over\partial\omega}\Biggr)_{\omega=\varepsilon_1+\varepsilon_2}
f_1f_2(1-f_3-f_4)
\nonumber\\
&\times
\left|t_{\rm sc}\!\Bigl(\!\varepsilon_1\!+\!\varepsilon_2\!+\!\Delta_E,
\V k\!+\!{\V \Delta_K\over 2},\!\V p+\!{\V \Delta_K\over 2},\V q,\V r\!+\!{\V \Delta_r},
t\!+\!{\Delta_t\over 2}\!\Bigr)\!\right|^2.
\label{Iout}
\end{align}
Thorough the paper all distribution functions and observables have the arguments
\begin{equation}
\begin{array}{rcl}
\varepsilon_1&\equiv&\varepsilon_a(\V k,\V r,t),\\
\varepsilon_2&\equiv&\varepsilon_b(\V p,\V r+\V \Delta_2,t),\\
\varepsilon_3&\equiv&\varepsilon_a(\V k-\V q+\V \Delta_K,\V r+\V \Delta_3,t+\Delta_t),\\
\varepsilon_4&\equiv&\varepsilon_b(\V p+\V q+\V \Delta_K,\V r+\V \Delta_4,t+\Delta_t)
\end{array}
\label{vc9dinv}
\end{equation}
and a bar indicates the reversed sign of $\Delta$s. If not otherwise noted all derivatives are explicit ones, i.e off-shell derivatives keeping the energy argument of $\Delta(\omega)$ as independent variable.

In the scattering-out (scattering-in is analogous) one can see the
distributions of quasiparticles $f_1f_2$ describing the probability of a given
initial state for the binary collision. The hole distributions describing the
probability that the requested final states are empty and the particle
distribution of stimulated collisions combine together in the final
state occupation factors like $1-f_3-f_4=(1-f_3)(1-f_4)+f_3f_4$.
The scattering rate covers the energy-conserving
$\delta$-function, and the differential cross section is given by the
modulus of the T-matrix reduced by the wave-function renormalizations $z_1\bar {z_2}\bar {z_3}\bar {z_4}$
\cite{K95}. 
We consider here the linear expansion in small scattering rates, therefore
the wave-function renormalization in the collision integral is of higher order.

All $\Delta$'s are derivatives of the scattering phase shift $\phi$,
\begin{equation}
t^R_{\rm sc}=\left|t_{\rm sc}\right|{\rm e}^{i\phi},
\label{vc3}
\end{equation}
according to the following list
\ba
&{\V \Delta_K}={1\over 2}{\partial\phi\over\partial \V r},
&\Delta_E&=-{1\over 2}{\partial\phi\over\partial t},
\qquad \Delta_t={\partial\phi\over\partial\omega},
\nonumber\\
&\V \Delta_2=
{\partial\phi\over\partial \V p}
-{\partial\phi\over\partial \V q}
-{\partial\phi\over\partial \V k},
&\V \Delta_3&=-{\partial\phi\over\partial \V k},
\nonumber\\
&{\V \Delta_4}=-{\partial\phi\over\partial \V q}-{\partial\phi\over\partial \V k},
&{\V \Delta_r}&={1\over 4}\left(\V \Delta_2\!+\!\V \Delta_3\!+\!{\V \Delta_4}\right)
.
\label{vc4g}
\end{align}

The quantum kinetic equation (\ref{vc17}) unifies the achievements of transport in
dense
gases with the
quantum transport of dense Fermi systems and was derived starting with the impurity problem \cite{SLMa96,SLM96b} and then for arbitrary Fermi systems \cite{SLM96,LSM97}. The
quasiparticle drift of Landau's
equation is connected with a dissipation governed by a nonlocal and non-instant
scattering integral in the spirit of Enskog corrections. These corrections
are expressed in terms of shifts in space and time that characterize
the non-locality of the scattering process \cite{MLSK99}. In this way quantum
transport was possible to recast into a quasi-classical picture suited for simulations.
The balance equations for the density, momentum and
energy include quasiparticle contributions and the correlated two-particle contributions beyond the Landau theory as we will demonstrate.

\begin{figure}[h]
  \includegraphics[width=4.2cm]{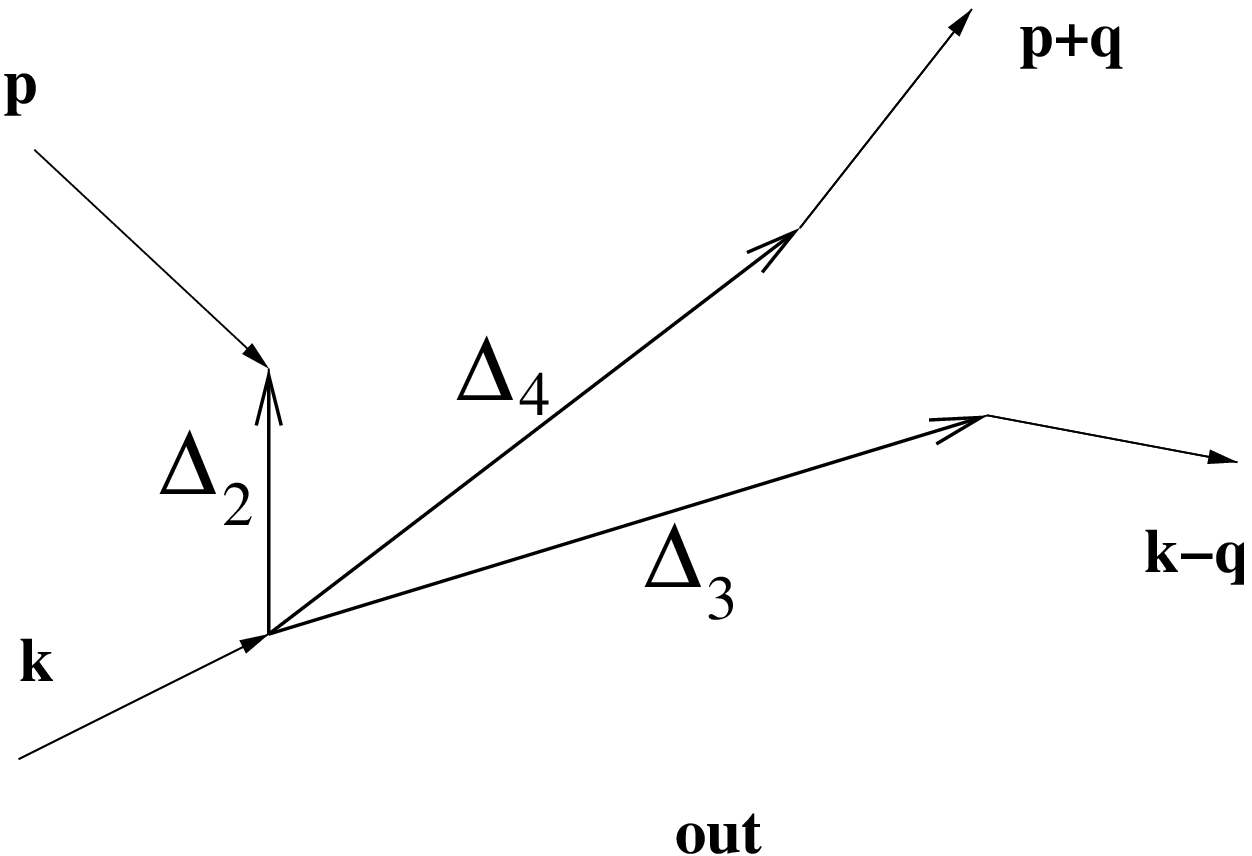}
  \includegraphics[width=4.2cm]{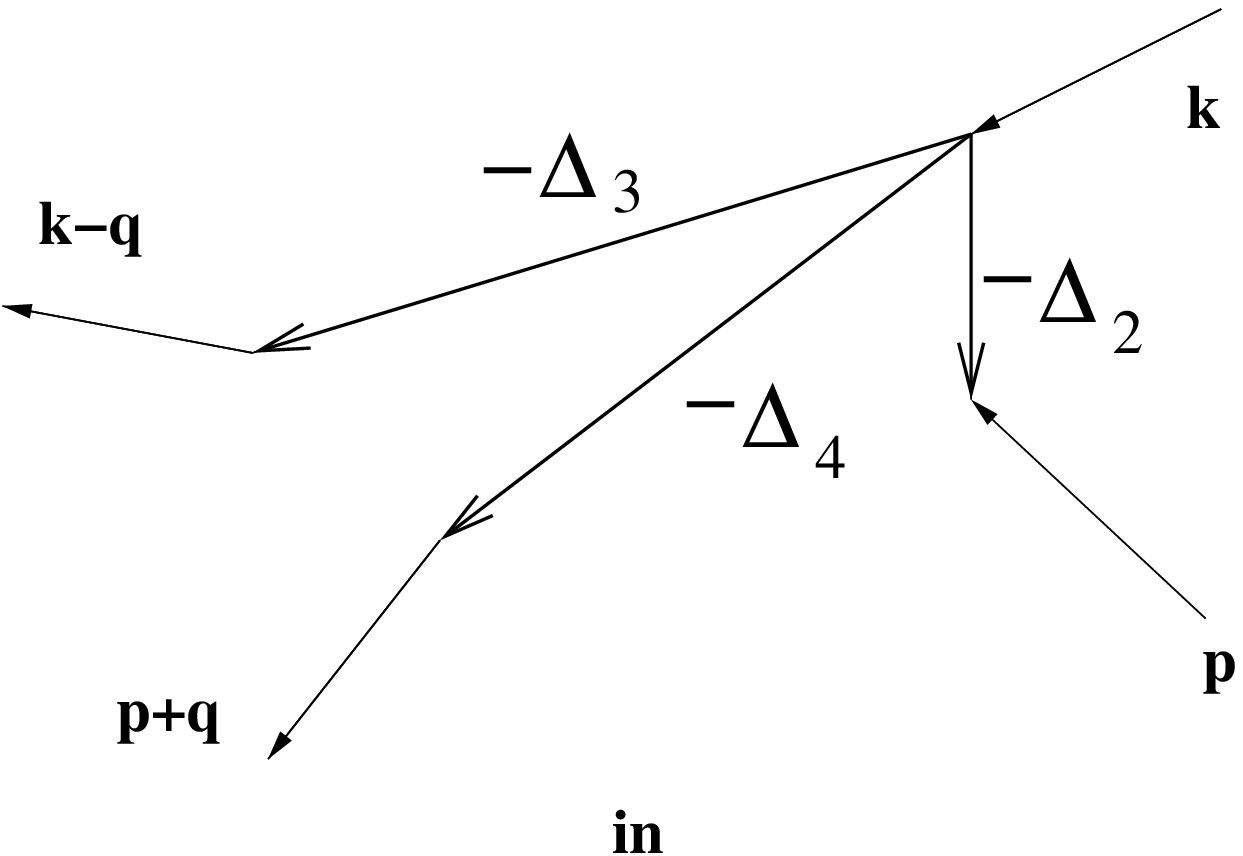}
\caption{Displacements in the effective collision of real particles.
In the scattering-out process, the momenta $k$ and $p$ correspond to the
initial states of particles $a$ and $b$, the momenta $k-q$ and $p+q$ to
the final states. In the scattering-in process, the picture is reversed.
}
\label{iif8}
\end{figure}

As special limits, this kinetic theory includes the Landau theory as well as the Beth-Uhlenbeck equation of state \cite{SRS90,MR94} which means correlated pairs.
The medium effects on binary collisions are shown to
mediate the latent heat which is the energy conversion between correlation
and thermal energy \cite{LSM99,LSM97}. In this respect the seemingly contradiction between particle-hole symmetry and time reversal symmetry in the collision integral was solved \cite{SLM98}.
Compared to the Boltzmann-equation, the presented form of virial
corrections only slightly increases the numerical demands in
implementations \cite{MSLKKN99,MT00,MTP00,MLNCCT01} since large cancellations
in the off-shell motion appear which are hidden usually
in non-Markovian behaviors. Details how to implement the nonlocal kinetic equation into existing Boltzmann codes can be found in \cite{MLNCCT01}.

Let us summarize the properties of the nonlocal kinetic equation
(\ref{vc17}). The drift is governed by the quasiparticle energy obtained
from the single-particle excitation spectrum. The scattering integral is
non-local and non-instant, including corrections to the conservation of
energy and momentum as it is illustrated in figure~\ref{iif8}. Neglecting the nonlocal shifts, the standard quasiparticle Boltzmann equation results with Pauli-blocking.

\section{Symmetries of collisions}
Integrating the kinetic equation it will be helpful to perform two transformations, once to interchange incoming and outgoing particles and once to exchange the collision partners $a$ and $b$.

\subsection{Transformation A}
\begin{figure}[h]
\parbox[]{9cm}{
\parbox[]{4cm}{
  \includegraphics[width=4cm]{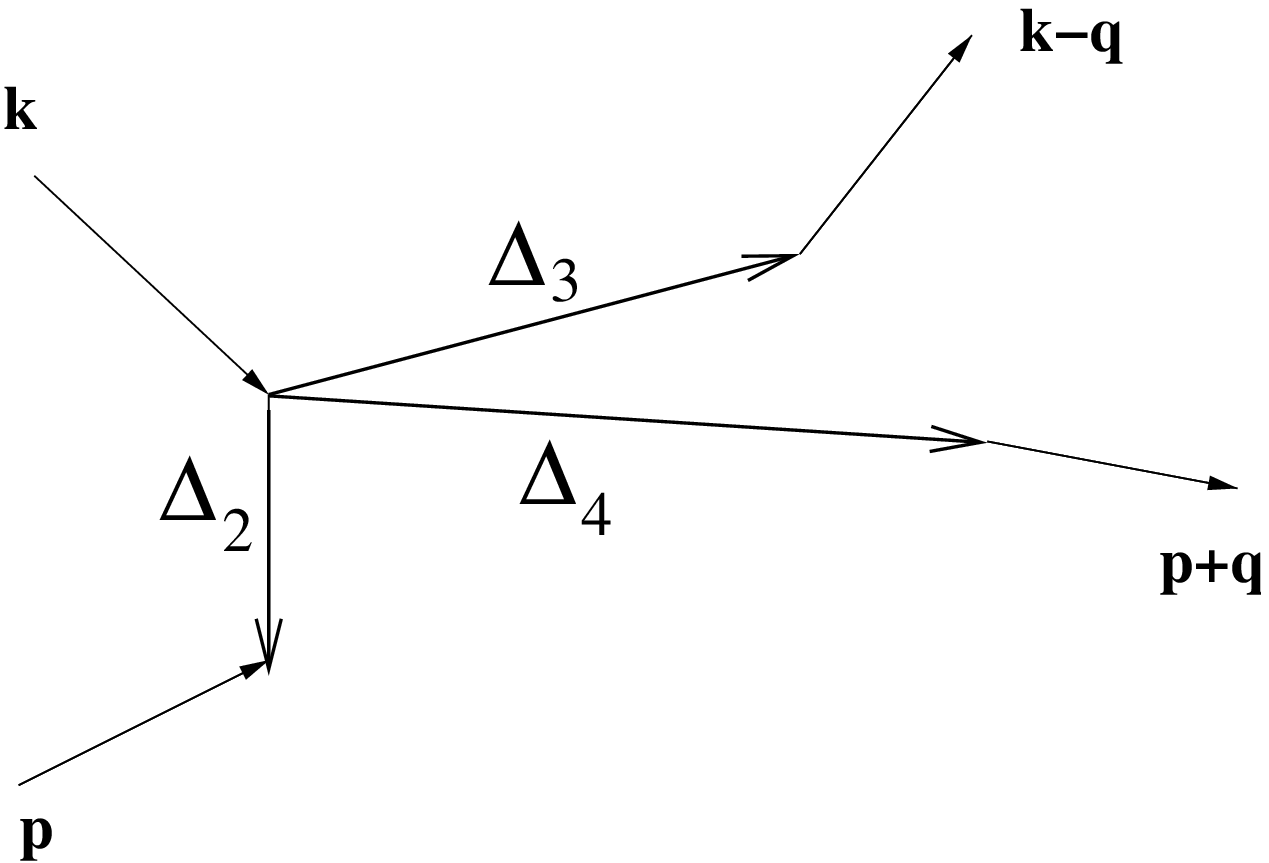}
}
\parbox[t]{0.5cm}{
\includegraphics[width=0.5cm]{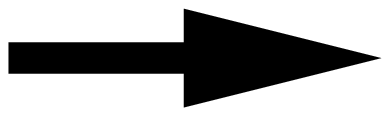}\\[2cm]
}
\parbox[]{4cm}{
  \includegraphics[width=4cm]{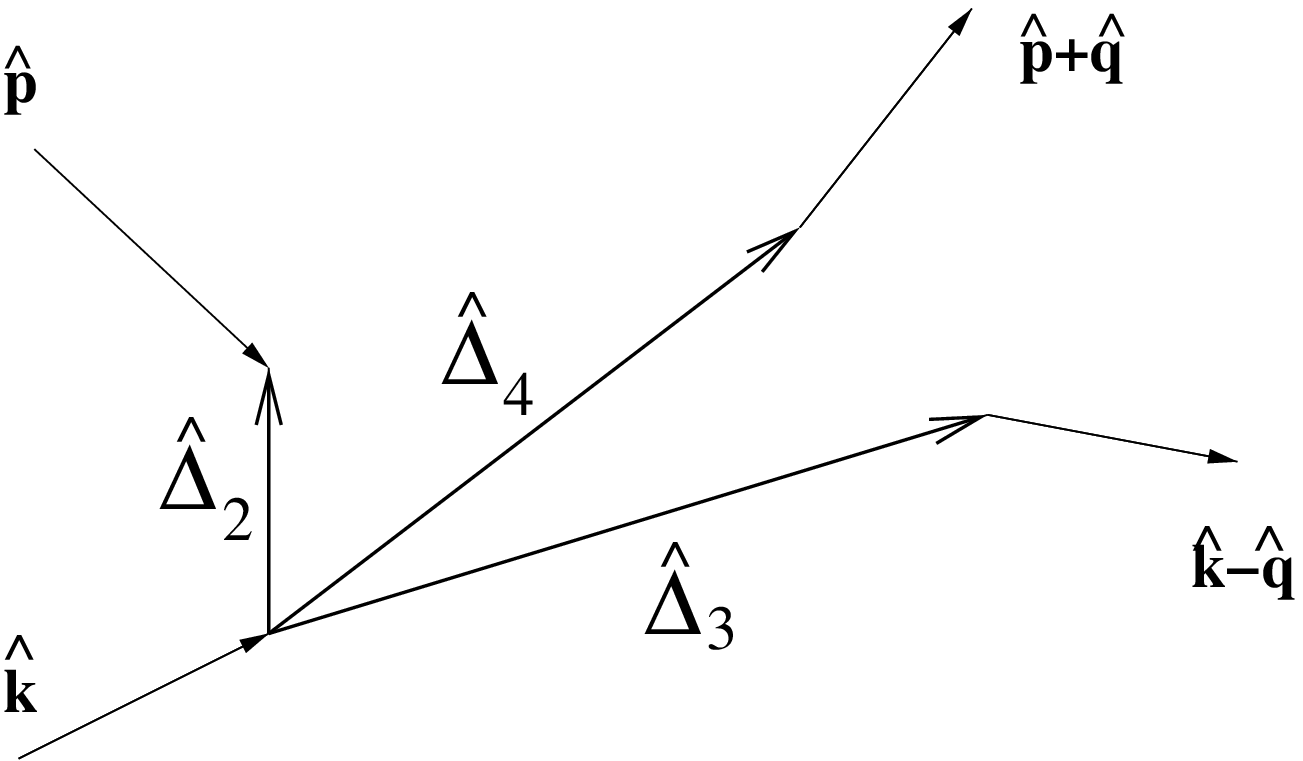}
}
}
\caption{Transformation A as interchange of particle $a$ and $b$ leading to (\ref{ec72}).}
\label{trafoA}
\end{figure}

The integrated kinetic equation (\ref{vc17})
is invariant if we
interchange particles $a$ and $b$ or labels $1\leftrightarrow 2$ and $3\leftrightarrow 4$. This is realized by the substitution
\be
a&=&\hat b,\quad
b=\hat a\nonumber\\
\V k&=&\hat {\V p},\quad
\V p=\hat {\V k},\quad
\V q=-\hat {\V q}.
\label{ec71}
\ee
The local T-matrix obeys this symmetry
\begin{equation}
t^R_{\rm sc}(\omega,\V k,\V p,\V q)=t^R_{\rm sc}(\omega,\V p,\V k,-\V q).
\end{equation}
However, this substitution changes the derivatives of the phase $\phi(\V k,\V p,\V q)=\hat \phi(\hat {\V k},\hat {\V p},\hat {\V q})=\hat \phi(\V p,\V k,-\V q)$ of the T-matrix (\ref{vc3}) as
\be
\frac{\partial \phi}{\partial{\V k}}=\frac{\partial \hat \phi}{\partial\hat {{\V p}}},\quad
\frac{\partial \phi}{\partial{\V p}}=\frac{\partial \hat \phi}{\partial\hat {{\V k}}},\quad
\frac{\partial \phi}{\partial{\V q}}=-\frac{\partial \hat \phi}{\partial\hat {{\V q}}}
\ee
leading to the relation between the displacements (\ref{vc4g})
\be
\V \Delta_2&=&-\hat {\V \Delta}_2\nonumber\\
\V \Delta_3&=&\hat {\V \Delta}_4-\hat {\V \Delta}_2\nonumber\\
{\V \Delta_4}&=&\hat {\V \Delta}_3-\hat {\V \Delta}_2\nonumber\\
{\V \Delta_r}&=&\hat {\V \Delta}_r-\hat {\V \Delta}_2.
\label{ec72}
\ee
Relations \eqref{ec72} merely show that the reference point has been
moved to the partner particle and shifts were correspondingly renamed,
see Fig.~\ref{trafoA}.

The $\Delta_t, \Delta_E$ and $\V {\Delta}_K$ remain
unchanged as well as the invariant combination
\ba
\V {\Delta}_{\mathrm{fl}}=\frac12\left(\V {\Delta}_3+\V {\Delta}_4-\V {\Delta}_2\right)=
\frac12\left(\hat {\V {\Delta}}_3+\hat {\V {\Delta}_4}-\hat {\V {\Delta}_2}\right)
\label{combi}
\end{align}
which is the distance between final and initial
geometrical centers of the colliding pair. It can be interpreted as the
distance on which particles travel during $\Delta_t$.

The quasiparticle energies (\ref{vc9dinv}) shift according to
\ba
\!&\varepsilon_1
\!&\rightarrow \, &\varepsilon_b(\V p,\V r,t)
\!&=\varepsilon_2(\V r\!-\!\V \Delta_2)\nonumber\\
\!&\varepsilon_2
\!&\rightarrow \, &\varepsilon_a(\V k,\V r\!-\!{\V \Delta_2},t)
\!&=\varepsilon_1(\V r\!-\!\V \Delta_2)\nonumber\\
\!&\varepsilon_3
\!&\rightarrow \,&\varepsilon_b(\V p\!+\!\V q\!+\!\V \Delta_K,\V r\!+\!{\V \Delta_4}\!-\!\V \Delta_2,t\!+\!\Delta_t)
\!&=\varepsilon_4(\V r\!-\!\V \Delta_2)\nonumber\\
\!&\varepsilon_4
\!&\rightarrow \,&\varepsilon_a(\V k\!-\!\V q\!+\!\V \Delta_K,\V r\!+\!{\V \Delta_3}\!-\!{\V \Delta_2},t\!+\!\Delta_t)
\!&=\varepsilon_3(\V r\!-\!\V \Delta_2)
\label{etrafoA}
\end{align}
where we denote only changes of the arguments of (\ref{vc9dinv}) explicitly.

\subsection{Transformation B}

\begin{figure}[h]
\parbox[]{9cm}{
\parbox[]{4cm}{
  \includegraphics[width=4cm]{trafoAa.eps}
}
\parbox[t]{0.5cm}{
\includegraphics[width=0.5cm]{arrow.eps}\\[2cm]
}
\parbox[]{4cm}{
  \includegraphics[width=4cm]{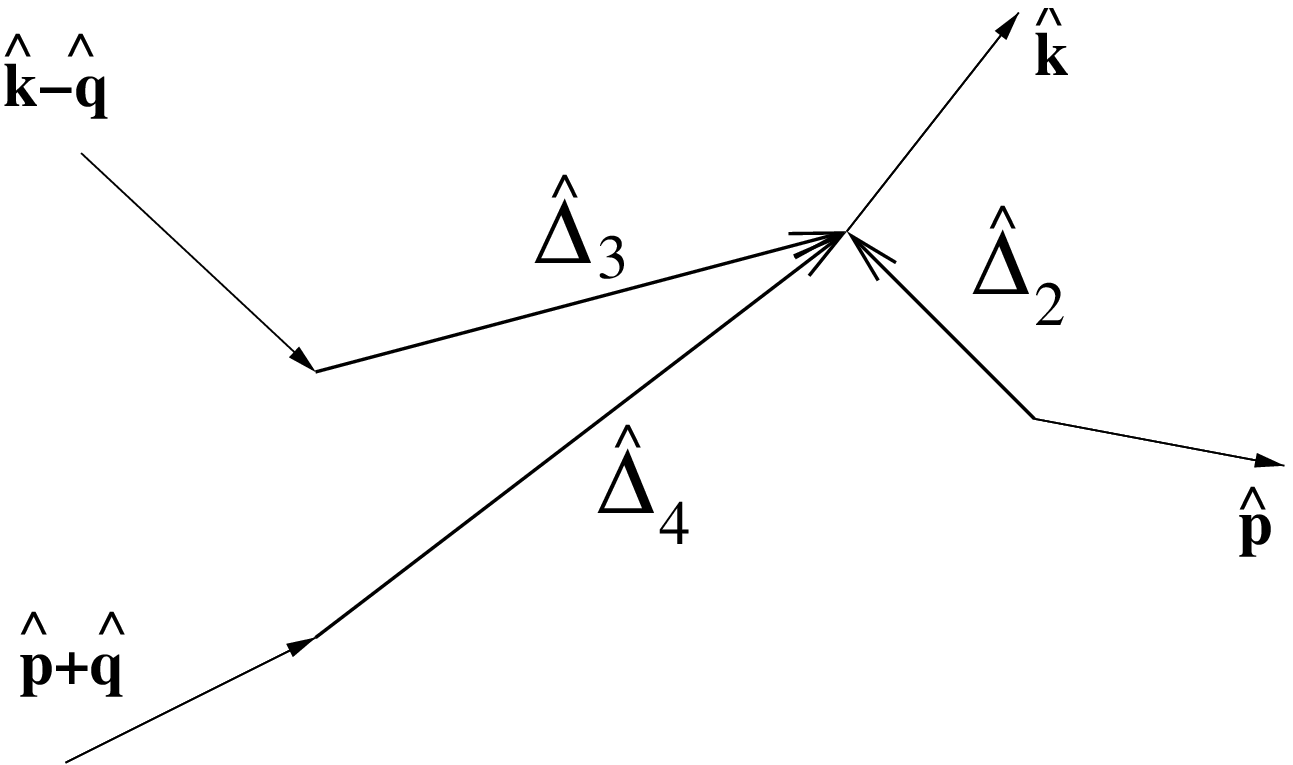}
}
}
\caption{Transformation B as interchange of incoming and outgoing particles leading to (\ref{ob54a}).}
\label{trafoB}
\end{figure}

The  interchange of initial and final states,
$1 \leftrightarrow  3$ and $2 \leftrightarrow 4$,
is accomplished by the substitution
\begin{equation}
\begin{array}{rcl}
\hat {\V k}&=&\V k-\V q,\quad
\hat {\V p}=\V p+\V q.\quad
\hat {\V q}=-\V q .
\end{array}
\label{ob29}
\end{equation}
The general symmetry of the
T-matrix with respect to the interchange of the initial and final
states,
\begin{equation}
t^R_{\rm sc}(\omega,\V k,\V p,\V q)=t^R_{\rm sc}(\omega,\V k-\V q,\V p+\V q,-\V q),
\label{ob30}
\end{equation}
implies that the differential cross section, the collision delay and the
energy gain do not change their forms by this substitution. All gradient corrections are explicitly in the form of
$\Delta$-corrections. 
Under this substitution, the
space displacements effectively behave as if we invert the collision,
\be
{\V \Delta_2}&\to&{\V \Delta_3}-{\V \Delta_4},\nonumber\\
{\V \Delta_4}&\to&{\V \Delta_3}-{\V \Delta_2},\nonumber\\
{\V \Delta_r}&\to&{\V \Delta_3}-{\V \Delta_r},
\label{ob54a}
\ee
while the other $\Delta$'s keeps their values and the combination (\ref{combi}) is invariant again. This is illustrated in figure~\ref{trafoB}.

The distributions of the in-scattering term can be translated into the ones of the out-scattering term if we interchange out- and in-going collisions which means to apply transformation B. Consequently, the arguments of the quasiparticle energies (\ref{vc9dinv}) transform as
\ba
&\varepsilon_1
&\rightarrow \, &\varepsilon_a(\V k-\V q,\V r,t)
&=\tilde \varepsilon_3\nonumber\\
&\bar\varepsilon_2
&\rightarrow \, &\varepsilon_b(\V p+\V q,\V r+{\V \Delta_4}-{\V \Delta_3},t)
&=\tilde \varepsilon_4\nonumber\\
&\bar \varepsilon_3
&\rightarrow \,&\varepsilon_a(\V k-\V \Delta_K,\V r-{\V \Delta_3},t-\Delta_t)
&=\tilde \varepsilon_1\nonumber\\
&\bar \varepsilon_4
&\rightarrow \,&\varepsilon_b(\V p-\V \Delta_K,\V r+{\V \Delta_2}-{\V \Delta_3},t-\Delta_t)
&=\tilde\varepsilon_2
\label{29}
\end{align}
where the $\tilde \varepsilon$ denotes the shift of momentum arguments by $-\V \Delta_K$, the spatial arguments by $-{\V \Delta_3}$, and the time arguments by $-\Delta_t$. If we transform the scattering-in (\ref{Iin}) with this transformation we obtain the order of distributions as the scattering-out (\ref{Iout}) however with these shifts in the different functions. Therefore we will abbreviate in the following
\begin{equation}
D_{in}=\left \{1\!-\!{\V \Delta_3}{\partial\over\partial \V r}\!-\!{\V \Delta_K}
\left({\partial\over\partial \V k}\!+\!{\partial\over\partial \V p}\right)\!-\!\Delta_t \pp t\right \}D,
\label{ec62}
\end{equation}
where we denote $D=D_{out}$ from (\ref{Iout}) with
\ba
D=\left|t_{\rm sc}\right|^22\pi\delta(\varepsilon_1\!+\!\varepsilon_2\!-\!
\varepsilon_3\!-\!\varepsilon_4\!+\!2\Delta_E)\left(1\!-\!f_3\!-\!f_4\right)f_1f_2.
\label{ec63}
\end{align} 
In case where this term $D$ will appear as prefactor to $\Delta$s we could ignore the shifts inside $D$ since our theory is linear in $\Delta$s. But consistently we will keep the shift as being the one of the out-scattering.

\subsection{Symmetrization of collision term}

In (\ref{vc17}), the differential cross section $|t_{\rm sc}|^2$ and 
the energy argument of the scattering phase shift is based on
initial states $\varepsilon_1+\varepsilon_2$. The
transformation B interchanges initial and final states and 
this sum energy becomes $\varepsilon_3+\varepsilon_4$. 
For a convenient implementation, we thus introduce the sum energy at the
center of the collision
\begin{equation}
\begin{array}{rcl}
E&=&{1\over 2}\left(\varepsilon_1+\varepsilon_2+\varepsilon_3+
\varepsilon_4\right),\\
\bar E&=&{1\over 2}\left(\varepsilon_1+\bar\varepsilon_2+
\bar\varepsilon_3+\bar\varepsilon_4\right).
\end{array}
\label{ob27}
\end{equation}

From the energy-conserving $\delta$-functions follows that on the energy
shell the centered energies equal to the arguments of the T-matrix,
\begin{equation}
\begin{array}{rcl}
\varepsilon_1+\varepsilon_2+\Delta_E&=&E,\\
\varepsilon_1+\bar\varepsilon_2-\Delta_E&=&\bar E.
\end{array}
\label{ob27a}
\end{equation}
The centered energies \eqref{ob27} are the physically
natural choice and we favor them against $\varepsilon_1+\varepsilon_2$
resulting from the quasiparticle approximation of Green's functions.
The centered energy argument, however, gives a non-trivial contribution
to the factor of the energy-conserving $\delta$ functions. This comes from the fact that any energy argument in the scattering-in term is given in terms of $\omega=\varepsilon_1+\bar \varepsilon_2$ while the scattering-out has the argument $\omega=\varepsilon_1+\varepsilon_2$.
Changing to the centered energies (\ref{ob27}) means 
mathematically
\ba
\delta[x\!-\!\Delta(x)]=\delta[x\!-\!\Delta(E)] \left [1\!+\!{\partial \Delta(x)\over \partial x}\left (\!1\!-\!{\partial E(x)\over \partial x}\right )\right ] 
\label{ob27c}
\end{align}
which one gets by comparing
$$
\delta[x-\Delta(x)]
=
{\delta(x-x_0) \over 1-{\partial \Delta(x)\over \partial x}}
\approx
\delta(x-x_0)
\left (1+{\partial \Delta(x)\over \partial x}\right )
$$
with
$$\delta[x\!-\!\Delta(E(x))]
\approx 
\delta(x\!-\!x_0) \left (1\!+\!{\partial \Delta(x)\over \partial x}{\partial E(x)\over \partial x}\right ) 
$$
in linear order of $\Delta$s.
 Using (\ref{ob27c}) with $x=\varepsilon_1+\bar \varepsilon_1$ 
and after substitution (\ref{ob27}) the $\delta$-function of the scattering-in 
reads 
\ba
&\left(\!\!1\!-\!{1\over 2}{\partial{\V \Delta_2}\over\partial \V r}\!-\!
{\partial\bar\varepsilon_2\over\partial \V r}
{\partial{\V \Delta_2}\over\partial\omega}\! \right )\!\! 
\left .\delta\bigl(\varepsilon_1\!+\!\bar\varepsilon_2
\!-\! \bar\varepsilon_3\!-\!
\bar\varepsilon_4\!-\!2\Delta_E\!\bigr)\!\right |_{\omega=\varepsilon_1\!+\!\bar\varepsilon_2}
\nonumber\\
&= \left[
1-{1\over 2}{\partial{\V \Delta_2}\over\partial \V r}-
{1\over 2}\left({\partial\bar\varepsilon_3\over\partial \V k}
+{\partial\bar\varepsilon_4\over\partial \V p}\right)
{\partial{\V \Delta_K}\over\partial\omega}
-{1\over 2}\left(
{\partial\bar\varepsilon_2\over\partial \V r}
{\partial{\V \Delta_2}\over\partial\omega}
\right .\right .\nonumber\\
&\left .\left .
+
{\partial\bar\varepsilon_3\over\partial \V r}
{\partial{\V \Delta_3}\over\partial\omega}+
{\partial\bar\varepsilon_4\over\partial \V r}
{\partial{\V \Delta_4}\over\partial\omega}\right)
 +{\partial\Delta_E\over\partial\omega}-{1\over 2}
{\partial\Delta_t\over\partial\omega}{\partial(\bar\varepsilon_3+
\bar\varepsilon_4)\over\partial t}
\right ]
\nonumber\\
&\times\left .\delta\left(\varepsilon_1+\bar\varepsilon_2-\bar\varepsilon_3-
\bar\varepsilon_4-2\Delta_E\right)\right |_{\omega=\bar E}
\label{ob28}
\end{align}
and the scattering-out is given by inverse signs of the $\Delta$s at $\omega=E$. Please remember that the quasiparticle energies and distributions have the shifts according to (\ref{vc9dinv}) and a bar indicates the reversed sign.

\section{Nonequilibrium thermodynamic properties}

\subsection{Local conservation laws}

Now we ask about the consequences of the kinetic equation 
to the thermodynamic properties. Far from equilibrium, the traditional thermodynamic 
quantities as the temperature and chemical potential do not capture 
the time-dependent properties of the system. Accordingly, we want to express the thermodynamic observables as functionals of the time-dependent quasiparticle distribution. Therefore we will multiply
the kinetic equation \eqref{vc17} with a variable $\xi_1=1,\V {k},\varepsilon_1,-k_{\mathrm B}\ln[f_1/(1-f_1)]$
and integrate over momentum. It results in the equation of continuity, the
Navier-Stokes equation, the energy balance and the evolution of the
entropy, respectively. 
All these conservation laws or balance equations for the mean thermodynamic observables 
\be
\langle \xi\rangle=\int{d^3k\over (2\pi)^3} \xi_1 f_1 
\ee 
will have the form
\be
{\partial \langle \xi \rangle \over \partial t}+\pp {\V r}  \V j_\xi={\cal I}_{\rm gain}.
\label{conserv}
\ee 
The additional gain on the right side might be due
to an energy or force feed from the outside or the  entropy production by 
collisions. The external potential is absorbed here in the quasiparticle energy and we can concentrate on the internal contributions due to correlations. Then we will obtain a gain only for the entropy while for density, momentum and energy the time change of the density $\langle \xi\rangle$  is exclusively  caused by
the divergence of the current $\V j_\xi$.
  

We will show from the balances of the kinetic equation that the particle 
density,
momentum flux (pressure), energy and entropy density consist of a quasiparticle part and a correlated contribution $\langle \xi \rangle =\xi^{\rm qp}+\xi^{\rm mol}$, respectively. The latter one takes the form of a molecular contribution as if two particles form a molecule. 
Proving the conservation laws (\ref{conserv}) and showing that also the currents consists of $\V j_\xi= \V j^{\rm qp}_\xi+\V j^{\rm mol}_\xi$ will be the ultimate goal to convince us about the consistency of the nonlocal kinetic equation.

\subsection{Drift contributions to balance equations}

The balance of quantity $\xi$ requires to evaluate the $\xi$-weighted momentum
integrals of the kinetic equation \eqref{vc17}, 
\be
\sum_a\!\int\!\frac{d^3 {k}}{(2\pi)^3}\xi_1{\partial f_1\over\partial t}
&+&\sum_a\!\int\!\frac{d^3 {k}}{(2\pi)^3}\xi_1\left({\partial\varepsilon_1\over\partial \V {k}}
{\partial f_1\over\partial \V {r}}
\!-\!{\partial\varepsilon_1\over\partial \V {r}}
{\partial f_1\over\partial \V {k}}\right)
\nonumber\\
&&=
\sum_a\int\frac{d^3 {k}}{(2\pi)^3}\xi_1\left(I_1^{\mathrm{in}}-I_1^{\mathrm{out}}\right).
\label{kineqweight}
\ee
The left-hand side includes terms which have been treated many times within the Boltzmann theory
and later extended to the Landau theory of Fermi liquids \cite{SMJ89,BaPe91}. Let us consider them first.

\subsubsection{Density balance from drift}

For the density $\xi_1=1$, the
left-hand side of the kinetic equation \eqref{kineqweight} gives
\ba
&
\int{d^3 {k}\over(2\pi)^3}
\left({\partial f_1\over\partial t}+{\partial\varepsilon_1\over\partial \V {k}}
{\partial f_1\over\partial \V {r}}-{\partial\varepsilon_1\over\partial \V {r}}
{\partial f_1\over\partial \V {k}}\right)
\nonumber\\&
={\partial\over\partial t}\int\!\!{d^3 {k}\over(2\pi)^3}f_1+
{\partial\over\partial \V {r}}\int\!\!{d^3 \! k\over(2\pi)^3}
{\partial\varepsilon\over\partial \V {k}}f_1
={\partial n_a^{\mathrm{qp}}\over\partial t}+
{\partial \V {j}_a^{\rm qp}\over\partial \V {r}}.
\label{ob51}
\end{align}
In Landau's theory the integration over the local collision integral of the Boltzmann equation is zero
and one finds that the divergence of the quasiparticle current
\begin{equation}
\V {j}_a^{\mathrm{qp}}=\int{d^3 {k}\over(2\pi)^3}
{\partial\varepsilon_1\over\partial \V {k}}f_1
\label{ob51a}
\end{equation}
and the time derivative of the quasiparticle density
\begin{equation}
n_a^{\mathrm{qp}}=\int{d^3 {k}\over(2\pi)^3}f_1
\label{ob5}
\end{equation}
sum to zero in the form of (\ref{conserv}).
The quasiparticle current includes the quasiparticle back flow \cite{LS94} which appears 
due to a non-symmetry of the quasiparticle energy $\varepsilon(k)\not=
\varepsilon(-k)$ resulting from a non-symmetry of the quasiparticle 
distributions.

\subsubsection{Energy}

We integrate now the kinetic equation (\ref{vc17}) multiplied with the energy
$\xi_1=\varepsilon_a(\V k,\V r,t)$. 
The drift side
\begin{eqnarray}
\sum\limits_a\int{d^3 \! k\over(2\pi)^3}\varepsilon_1 {\partial\over\partial t}f_1+{\partial\over\partial \V r}\sum\limits_a\int{d^3 \! k\over(2\pi)^3}
\varepsilon_1{\partial\varepsilon_1\over\partial \V k}f_1
\label{obe}
\end{eqnarray}
results in the divergence of the
quasiparticle energy current
\begin{equation}
\V {j}^{\rm qp}_E=\sum\limits_a\int{d^3 \! k\over(2\pi)^3}
\varepsilon_1{\partial\varepsilon_1\over\partial \V k}f_1
\label{ob51e}
\end{equation}
and the first term of (\ref{obe})
\be
\sum\limits_a\int{d^3 \! k\over(2\pi)^3}\varepsilon_1 {\partial\over\partial t}f_1.
\label{ec2}
\ee

When $\Delta$'s tend to zero the collision integral vanishes after integration over $\varepsilon$ and the energy balance is (\ref{obe}). Obviously (\ref{ec2})
  has to be rearranged into the time
derivative. 
In the absence of non-local collisions which corresponds to Landau's concept of quasiparticles, the quasiparticle energy
$\varepsilon$ equals the functional derivative of the energy density,
\begin{equation}
\varepsilon={\delta{\cal E}^{\Delta=0}\over\delta f}.
\label{ec2d}
\end{equation}
With the help of (\ref{ec2d}) 
the drift term (\ref{obe}) attains the desired form,
\begin{equation}
\sum\limits_a\int{d^3 \! k\over(2\pi)^3}\varepsilon_1 {\partial\over\partial t}f_1=\sum_a\int{d^3 \! k\over(2\pi)^3}
{\delta{\cal E}^{\Delta=0}\over\delta f_1}{\partial f_1\over\partial t}=
{\partial{\cal E}^{\Delta=0}\over\partial t}.
\label{ec2ba}
\end{equation}

Landau's functional relation (\ref{ec2d}) is consistent with the 
Boltzmann equation and is particularly useful for phenomenological
quasiparticle energies \cite{L56,BaPe91}. 
The variational energy (\ref{ec2d}) makes the conservation laws very
convenient and results in correct collective motion. Here
we use the quasiparticle energy identified as the pole of
the Green's function. Except for simple approximations,
these two definitions lead to different values of quasiparticle
energies.  In the theory of liquid $^3_2$He, the difference 
between these two definitions is know as the rearrangement energy 
\cite{GH83}. 
A relation between these quasiparticle energies and the rearrangement
energy has been discussed in \cite{LSM99}.

The simplicity of Landau's variational approach makes
his concept of quasiparticles very attractive. On the
other hand, the Green's function pole represents the true
dispersion law of single-particle excitation, therefore the
pole definition leads to a better description of the local
distribution of particles. Of course, it is on cost of more
complex balance equations.
The quasiparticle contributions for all thermodynamical
quantities we discuss are complete if we evaluate
the collision contributions.

\subsubsection{Balance of forces}

For the momentum balance
one multiplies the kinetic equation (\ref{vc17}) with the $j$-th component of
momentum $\V k$, i.e. $\xi=k_j$, and integrates
over momentum $k$. The ${\partial f\over\partial t}$ results in the time
derivative of the momentum density of quasiparticles
\begin{equation}
{\cal Q}_j^{\rm qp}=\sum_a\int{d^3 \! k\over(2\pi)^3}k_jf_1.
\label{ec51ab}
\end{equation}
The other parts of the drift side can be rearranged by integration by parts as
\ba
&\sum_a\int{d^3 \! k\over(2\pi)^3}k_j\left(
{\partial\varepsilon_1\over\partial \V k}{\partial f_1\over\partial \V r}-
{\partial\varepsilon_1\over\partial \V r}{\partial f_1\over\partial \V k}\right)
\nonumber\\
&=\!\sum_{i,a}\!{\partial\over\partial r_i}
\!\int\!\!{d^3 \! k\over(2\pi)^3}\!\left(\! k_j
{\partial\varepsilon_1\over\partial k_i}\!+\!\delta_{ij}\varepsilon_1\!\right)f_1
\!-\!
\sum_a\!\!\int\!\!{d^3 \! k\over(2\pi)^3}\varepsilon_1{\partial f_1\over\partial r_j}.
\label{ec52}
\end{align}
Eq. (\ref{ec2d}) allows to write the last term of (\ref{ec52}) as
the gradient of the energy density,
\begin{equation}
{\partial{\cal E}^{\Delta=0}\over\partial \V r_j}=
\sum_a\int{d^3 \! k\over(2\pi)^3}{\delta{\cal E}^{\Delta=0}\over\delta f}
{\partial f\over\partial r_j}\equiv
\sum_a\int{d^3 \! k\over(2\pi)^3}\varepsilon{\partial f\over\partial r_j}.
\label{ec53}
\end{equation}
In such a way (\ref{ec52}) becomes the 
quasiparticle stress tensor \cite{SMJ89,BaPe91},
\begin{equation}
{\Pi}_{ij}^{\Delta=0}=\sum_a\int{d^3 \! k\over(2\pi)^3}\left(k_j
{\partial\varepsilon_1\over\partial k_i}+\delta_{ij}\varepsilon\right)f_1-
\delta_{ij}{\cal E}^{\Delta=0}.
\label{ec55}
\end{equation}
The quasiparticle momentum-force balance from the drift becomes therefore
\begin{align}
&\sum_a\int{d^3 {k}\over(2\pi)^3}k_j\left({\partial f_1\over\partial t}+
{\partial\varepsilon_1\over\partial \V {k}}
{\partial f_1\over\partial \V {r}}-{\partial\varepsilon_1\over\partial \V {r}}
{\partial f_1\over\partial \V {k}}\right)
\nonumber\\
&~~={\partial {\cal Q}_j^{\mathrm{qp}}\over\partial t}+
\sum_i{\partial {\Pi}_{ij}^{\mathrm{\Delta=0}}\over\partial r_i},
\label{mfbal}
\end{align}
as the one for the local Boltzmann equation or Landau's theory since without
shifts the collision integral vanishes due to momentum conservation. We will
obtain additional contributions from the nonlocal collision integral.

\subsubsection{Entropy}

Finally, the single-particle entropy  density distribution is given by \cite{L56}
\be
s_a(\V k,\V r,t)=-k_B\left [ f_1 \ln f_1+(1-f_1) \ln (1-f_1) \right ]
\label{singentrop}
\ee
which is the generalization of the classical expression towards quantum effects including the Pauli-blocking. The first sum in (\ref{singentrop}) is the entropy of particles but with the quantum quasiparticle distribution. The second sum one can consider as the entropy of holes $1-f$ as if they are just a second sort of particles. 

Since any derivative of (\ref{singentrop}) leads to the derivative of the
distribution $\partial s_1=-k_B \ln[f_1/(1-f_1)]\partial f_1$ it is advisable to
multiply the kinetic equation (\ref{vc17}) with $\xi_1=-k_B \ln [f_1/(1-f_1)]$
and to integrate over $\V k$.
The drift side becomes
\ba
&\sum\limits_a\int{d^3 \! k\over(2\pi)^3}{\partial s_1\over\partial t}+\sum\limits_a\int{d^3 \! k\over(2\pi)^3}\left [{\partial \varepsilon_1 \over\partial \V k}{\partial s_1 \over\partial \V r}-{\partial \varepsilon_1 \over\partial \V r}{\partial s_1 \over\partial \V k}\right ]
\nonumber\\&={\partial {\cal S}^{\rm qp} \over\partial t}+{\partial \V {j}^{\rm qp}_S\over\partial \V r}.
\label{obent}
\end{align}
It results into the divergence of the
quasiparticle entropy current
\ba
&\V {j}^{\rm qp}_S(\V r,t)=\sum\limits_a\int{d^3 \! k\over(2\pi)^3}
{\partial\varepsilon_1\over\partial \V k}s_a(\V k,\V r,t)
\nonumber\\
&=-k_B \sum\limits_a\int{d^3 \! k\over(2\pi)^3}
{\partial\varepsilon_1\over\partial \V k}
\left [ f_1 \ln f_1+(1-f_1) \ln (1-f_1) \right ]
\label{ob51ent}
\end{align}
and the time derivative of the quasiparticle entropy 
\ba
&{\cal S}^{\rm qp}(\V r,t)=\sum\limits_a\int{d^3 \! k\over(2\pi)^3} s_a(\V k,\V r,t)
\nonumber\\
&=-k_B \sum\limits_a\int{d^3 \! k\over(2\pi)^3} \left [ f_1 \ln f_1+(1-f_1) \ln (1-f_1) \right ]
\label{entropquasi}
\end{align}
as integral over (\ref{singentrop}). The arguments of $f_1$, $f_2$ etc. follow the notation of (\ref{vc9dinv}).

For the entropy balance, the collision integral does not vanish even neglecting shifts providing an explicit entropy gain. 
The interesting question is how the molecular part of entropy will look like, what balance we get and whether we can prove Boltzmann's H-theorem, i.e. the second law of thermodynamics. If we manage to derive the expressions including shifts and to prove the H-theorem, this includes, of course, then also the simpler case for local Boltzmann equation neglecting shifts.

\subsection{Molecular contributions to observables from collision integral}

Besides the known quasiparticle contributions to the observables of the last chapter, there appear explicit binary correlations due to the nonlocal collision integral. The remaining parts of this chapter presents a new systematic way to derive these correlated observables.

\subsubsection{Expansion properties}

Now we search for the terms arising from the nonlocal collision integral  (\ref{vc17}). 
Multiplying the latter one with $\xi_1$, integrating and applying the B-transform to the in-scattering part, we obtain from (\ref{ob28}) the structure 
\ba
&\int\!{d^3 \! kd^3 \! pd^3 \! q\over(2\pi)^9}
\xi_3(\V k\!-\!\V \Delta_K,r\!- \!{\V \Delta_3},t\!-\!\Delta_t) [1+{\cal T}_{in}]
\nonumber\\
&~~~\times\left [1\!-\!{\V \Delta_3}{\partial\over\partial \V r}\!-\!{\V \Delta_K}
\left({\partial\over\partial \V k}\!+\!{\partial\over\partial \V p}\right)-\Delta_t \pp t\right ] D
\nonumber\\
&-\int\!{d^3 \! kd^3 \! pd^3 \! q\over(2\pi)^9}\xi_1 [1+{\cal T}_{out}] D
\label{balance}
\end{align}
where we abbreviated (\ref{ec63}) and adopt the notation (\ref{vc9dinv}) of the arguments for the observable $\xi$. One has (\ref{29}) when B-transforming and in this notation $\xi_3(\V k\!-\!\V \Delta_K,r\!- \!{\V \Delta_3},t\!-\!\Delta_t)=\xi_a(\V k-\V q,\V r,t)$.
The factors (\ref{ob28}) transform into
\ba
&{\cal T}_{in}=
-{1\over 2}{\partial({\V \Delta_3}-{\V \Delta_4})\over\partial \V r}-
{1\over 2}\left({\partial\tilde\varepsilon_1\over\partial \V k}
+{\partial\tilde\varepsilon_2\over\partial \V p}\right)
{\partial{\V \Delta_K}\over\partial\omega}
\nonumber\\&
-{1\over 2}\left(
{\partial\tilde\varepsilon_4\over\partial \V r}
{\partial({\V \Delta_3}-{\V \Delta_4})\over\partial\omega}
+
{\partial\tilde\varepsilon_1\over\partial \V r}
{\partial{\V \Delta_3}\over\partial\omega}
+
{\partial\tilde\varepsilon_2\over\partial \V r}
{\partial({\V \Delta_3}-{\V \Delta_2})\over\partial\omega}\right)
\nonumber\\
&+ 
{\partial\Delta_E\over\partial\omega}-{1\over 2}
{\partial\Delta_t\over\partial\omega}{\partial(\tilde\varepsilon_1+
\tilde\varepsilon_2)\over\partial t}
\nonumber\\
&{\cal T}_{out}=
{1\over 2}{\partial{\V \Delta_2}\over\partial \V r}\!+\!
{1\over 2}\left({\partial\tilde\varepsilon_3\over\partial \V k}
\!+\!{\partial\tilde\varepsilon_4\over\partial \V p}\right)
{\partial{\V \Delta_K}\over\partial\omega}
\nonumber\\
&+
{1\over 2}\left(
{\partial\tilde\varepsilon_2\over\partial \V r}
{\partial{\V \Delta_2}\over\partial\omega}
+
{\partial\tilde\varepsilon_3\over\partial \V r}
{\partial{\V \Delta_3}\over\partial\omega}+
{\partial\tilde\varepsilon_4\over\partial \V r}
{\partial{\V \Delta_4}\over\partial\omega}\right)
\nonumber\\
&- 
{\partial\Delta_E\over\partial\omega}+{1\over 2}
{\partial\Delta_t\over\partial\omega}{\partial(\tilde\varepsilon_3+
\tilde\varepsilon_4)\over\partial t}
\label{inout}
\end{align}
where the unchanged out-scattering one is just (\ref{ob28}) with reversed
signs and the in-scattering one appears since we have applied transformation B
to (\ref{ob28}). Since our theory is linear in $\Delta$s we ignore the shifts inside $\varepsilon_i$, i.e. we can use $\tilde \varepsilon=\varepsilon=\bar\varepsilon$ when they appear as factors with $\Delta$s.

To start with the treatment of all the following expansions it is very helpful to observe that we can consider the arguments of $\Delta$s before expansion either being $E=\epsilon_1+\epsilon_2+o(\Delta_E)$ or alternatively $E'=\epsilon_3+\epsilon_4+o(\Delta_E)$ up to first order in $\Delta$s due to the energy conservation in $D$. To see this we expanding the equality $D\Delta(E)=D\Delta(E')$ up to first order for any $\Delta$ 
\be
D_0+\partial D \Delta (E)=D_0+\partial D \Delta (E')
\ee
with the corresponding derivative $\partial$.
From this we now subtract the equality $0=\partial (D\Delta(E))=\partial (D\Delta(E'))$ due to the $\delta$-function, to get the relation
\be
D \partial \Delta(E)=D \partial \Delta(E')
\label{relt}
\ee
which we will use later.

\subsubsection{Correlated observables}

In order to make the different parts transparent we concentrate successively on specific terms and collect them together in the end.

For the time derivative and $\Delta_t$ terms in (\ref{balance}) and (\ref{inout}),
we employ transformation A, add with the original expression and divide by 2 resulting into the terms under integration
\be
&&{\xi_3+\xi_4-\xi_1-\xi_2\over 2} D-{\Delta_t\over 2}\pp t \left [(\xi_3+\xi_4) D \right ]
\nonumber\\
&&-{\xi_3+\xi_4\over 4} 
D\left (
{\partial \Delta_t\over \partial \omega}{\partial (\varepsilon_1+\varepsilon_2)\over \partial t}
-2{\partial\Delta_E\over \partial \omega}
\right )
\nonumber\\&&
-{\xi_1+\xi_2\over 4}D \left ( 
{\partial \Delta_t\over \partial \omega}{\partial (\varepsilon_3+\varepsilon_4)\over \partial t}
-2{\partial\Delta_E\over \partial \omega}
\right ).
\label{43}
\ee
The terms in the brackets form the total (on-shell) derivative as follows.
From the definition (\ref{vc4g}) we have the identity 
$\p \omega \Delta_E(t,\omega)=-\p t \Delta _t(t,\omega)/2$ and therefore for any argument $x(t)$
\ba
{\partial\Delta_t(t,\omega)\over\partial \omega}{\partial x \over\partial t}\!-\!2{\partial\Delta_E\over\partial\omega}
&=
{\partial\Delta_t(t,\omega)\over\partial\omega}{\partial x \over\partial t}
+{\partial\Delta_t(t,\omega)\over\partial t}
\nonumber\\&
={\partial^{on}\over \partial t}\Delta_t (t,x).
\label{onshell}
\end{align}
This means we can write for (\ref{43})
\ba
&{\xi_3+\xi_4-\xi_1-\xi_2\over 2} D-{\Delta_t(E)\over 2}{\pp t}^{on} \left [(\xi_3+\xi_4) D \right ]
\nonumber\\
&-{\xi_3+\xi_4\over 4} 
D{\partial \over \partial t}^{on}\Delta_t(E)
-{\xi_1+\xi_2\over 4}
D{\partial \over \partial t}^{on}\Delta_t(E')
\label{43a}
\end{align}
Using (\ref{relt}) we can add the last two expression,
\be
&&
{\xi_3+\xi_4-\xi_1-\xi_2\over 2} D-{\Delta_t(E) \over 2 } {\partial \over \partial t}^{on}[D(\xi_3+\xi_4)]
\nonumber\\&&-{\xi_3+\xi_4+\xi_1+\xi_2\over 4} 
D{\partial \over \partial t}^{on}\Delta_t(E)
\nonumber\\&&
={\xi_3\!+\!\xi_4\!-\!\xi_1\!-\!\xi_2\over 2} D
-{\partial \over \partial t}^{on}\left [{\xi_3\!+\!\xi_4\!\over 2} 
D \Delta_t(E)\right ]
\nonumber\\&&
-{\partial^{on} {\Delta_t(E)} \over \partial t}  \left [{\xi_3+\xi_4-\xi_1-\xi_2 \over 4}D\right  ]
\label{ec8ent}
\ee
such that we obtain finally
\ba
&-\pp t \frac{1}{2} \sum_{ab}\int{d^3 \! kd^3 \! pd^3 \! q\over(2\pi)^9}\Delta_t D (\xi_3+\xi_4)
\nonumber\\&
+\frac{1}{2} \sum_{ab}\int{d^3 \! kd^3 \! pd^3 \! q\over(2\pi)^9} \left (1+{1\over 2}\pp t\Delta_t\right )D (\xi_3+\xi_4-\xi_1-\xi_2).
\label{entc}
\end{align}
The first term is the negative of the time derivative of a molecular contribution to the observable
\ba
{\xi}^{\rm mol}={1\over 4}\sum_{ab}\int{d^3 \! kd^3 \! pd^3 \! q\over(2\pi)^9}
D
\Delta_t
(\xi_3+\xi_4)
\label{phicurr}
\end{align}
which will be added to the quasiparticle part from the drift side.
It possesses a form which can be statistically understood. With the rate 
$D$ of (\ref{ec63}) molecules are formed and multiplied with their lifetime $\Delta_t$ to provide the probability with which the observables $\xi$ occur in the molecular state.

The observable gain is the  second part of (\ref{entc}),
\ba
&{\cal I}_{\rm gain}^\xi
=
\frac{1}{2} \sum_{ab}\!\int\!\!{d^3 \! kd^3 \! pd^3 \! q\over(2\pi)^9} \!\left (\!1\!+\!{1\over 2}\pp t\Delta_t\!\right )\!D(\xi_3\!+\!\xi_4\!-\!\xi_1\!-\!\xi_2).
\label{ec10ent}
\end{align}
We see that for density $\xi=1$ we do not have a gain. For momentum gain
$\xi=k_j$ we get from (\ref{ec10ent}) linear in $\Delta$
\begin{equation}
{\cal F}^{\rm gain}_j=\sum_{ab}\int{d^3 \! kd^3 \! pd^3 \! q\over(2\pi)^9}
D
\Delta_{Kj}
.
\label{ec56fg}
\end{equation}
Dividing and multiplying by $\Delta_t$ under the integral we see that the
momentum gain is the probability $D\Delta_t$ to form a molecule multiplied with the force $\V \Delta_K/\Delta_t$ exercised during the delay time $\Delta_t$ from the environment by all other particles.
This momentum gain
(\ref{ec56fg}) can be exactly recast together with the last term of the drift
(\ref{ec52}) into a spatial derivative 
\be
\sum_a\int{d^3 \! k\over(2\pi)^3}\varepsilon{\partial f\over\partial \V r_j}
+{\cal F}^{\rm gain}_j={\partial{\cal E}^{\rm qp}\over\partial \V r_j}
\label{ec79}
\ee
of the quasiparticle energy functional \cite{LSM97}
\ba
{\cal E}^{\rm qp}&=\sum_a\int{d^3 \! k\over(2\pi)^3}f_a(k){k^2\over 2m}
\nonumber\\+&
{1\over 2}\sum_{ab}\int{d^3 \! kd^3 \! p\over(2\pi)^6}f_a(k)f_b(p)
{t}_{\rm sc}(\varepsilon_1+\varepsilon_2,k,p,0)
\label{ec22cb}
\end{align}
instead of the Landau functional (\ref{ec53})
which was valid only in local approximation.

For the energy gain $\xi=\varepsilon$ we get from (\ref{ec10ent})
\begin{equation}
{\cal I}_{\rm gain}^E=\sum_{ab}\int{d^3 \! kd^3 \! pd^3 \! q\over(2\pi)^9}
D
\Delta_E.
\label{ec10}
\end{equation}
It represents the mean power $\Delta_E/\Delta_t$ exerted on the collision multiplied with the probability to form a molecule $D\Delta_t$.
As proved in \cite{LSM97}
this energy gain combines together with the first term of (\ref{obe}) into the total time derivative of the quasiparticle energy functional (\ref{ec22cb})
\begin{equation}
\sum_a\int{d^3 \! k\over(2\pi)^3}\varepsilon
{\partial f\over\partial t}
-{\cal I}_{\rm gain}^E=
{\partial{\cal E}^{\rm qp}\over\partial t}.
\label{ec11}
\end{equation}

The entropy gain (\ref{ec10ent}) with $\xi_1=-k_B \ln f_1/(1-f_1)$
we will discuss later in chapter~\ref{agains}.

Summarizing so far, we have arrived  from the time parts of the collision integral at the balance equation almost at the form (\ref{conserv})
\be
{\partial (\xi^{\rm qp}+\xi^{\rm mol})\over \partial t}+{\partial \V j_\xi^{\rm qp}\over \partial \V r}={\cal I}_{\rm space}
\label{balanceEa}
\ee
where it remains to show that the spatial and momentum derivatives of (\ref{balance}), indicated as 
${\cal I}_{\rm space}$, can be written as the divergence of the molecular energy current.

\subsection{Molecular current contributions from collision integral}

First we need a guide how the corresponding molecular currents will look
like. Most simply this is seen from the spatial gradients of $D$ in
(\ref{balance}) and (\ref{inout})
\ba
\xi_3 D(\V r\!-\!{\V \Delta_3})\!-\!\xi_1 D(\V r)=(\xi_3\!-\!\xi_1) D\!-\!{\V \Delta_3} \xi_3 \pp {\V r} D\!+\!o(\Delta^2).
\label{d1}
\end{align}
Alternatively we can apply first the transformation A and then expand
\ba
&\xi_4 D(\V r-{\V \Delta_2}-({\V \Delta_4}-{\V \Delta_2}))-\xi_2 D(\V r-{\V \Delta_2})
\nonumber\\&
=(\xi_4-\xi_2) D-{\V \Delta_4} \xi_4 \pp {\V r} D+{\V \Delta_2}\xi_2 \pp {\V r} D+o(\Delta^2).
\label{d2}
\end{align}
The shifts inside the $\xi$s can be neglected since we consider only linear
orders. We add (\ref{d1}) and (\ref{d2}) and divide by 2 to get
besides the already counted gain term (\ref{ec10ent})  
the first-order gradient term
\be
{\cal I}_{\rm space}^D=
\frac 1 2 ({\V \Delta_2}\xi_2 -{\V \Delta_3} \xi_3 -{\V \Delta_4} \xi_4) \pp {\V r} D.
\label{s1}
\ee
This suggests how the molecular observable current $\V j_\xi$
will look like provided we find the remaining terms such that (\ref{s1})
becomes the divergence of the molecular current, $\p {\V r} \V j_\xi^{\rm mol}$. 

We consider now the frequency derivatives of the spatial shifts in
(\ref{balance}) and (\ref{inout}), again adding the A-transform expression and 
dividing by 2. Collecting them together one gets 
\ba
&\frac D 2 {\partial E\over \partial \V r} \left ( 
\xi_2 {\partial {\V \Delta_2}\over \partial \omega}
-\xi_3 {\partial {\V \Delta_3}\over \partial \omega}
-\xi_4 {\partial {\V \Delta_4}\over \partial \omega} \right )
\nonumber\\&
\!+\!{D\over 4}(\xi_3\!+\!\xi_4\!-\!\xi_1\!-\!\xi_2)\left ( 
{\partial \varepsilon_2 \over \partial \V r} {\partial {\V \Delta_2}\over \partial \omega}
\!+\!{\partial \varepsilon_3 \over \partial \V r}{\partial {\V \Delta_3}\over \partial \omega}
\!+\!{\partial \varepsilon_4 \over \partial \V r}{\partial {\V \Delta_4}\over \partial
  \omega} \right )
\nonumber\\& 
={\cal I}_{\rm space}^\omega+{\cal I}_{\rm gain}^\omega.
\label{s2}
\end{align}
The first part fits the derivative in (\ref{s1}) while the second part obviously
counts together with the second part of (\ref{entc}), i.e. it is  
a gain term (\ref{ec10ent}) due to the factor $(\xi_3+\xi_4-\xi_1-\xi_2)$.

Next, we collect the spatial derivatives of $\xi$ and the spatial shifts of
(\ref{balance}) with (\ref{inout}). When applying the A-transform, $\xi_3(r-\Delta_3)\to \xi_4(r-\Delta_2-(\Delta_4-\Delta_2))$, one gets
\ba
&-{\V \Delta_3} \pp {\V r} \xi_3-{\xi_3 \over 2} \pp {\V r} ({\V \Delta_3}-{\V \Delta_4})-{\xi_1\over 2}\pp {\V r} {\V \Delta_2}
\nonumber\\&=
- {{\V \Delta_4}} \pp {\V r} \xi_4+{\xi_4\over 2} \pp {\V r} ({\V \Delta_3}-{\V \Delta_4})+ {{\V \Delta_2}\over 2} \pp {\V r} \xi_2
\nonumber\\&
=
-{{\V \Delta_3}\over 2}\pp {\V r}\xi_3 -{{\V \Delta_4}\over 2}\pp {\V r}\xi_4+{{\V \Delta_2}\over 2}\pp {\V r}\xi_2  
\nonumber\\&
+{\xi_2-\xi_1\over 4}\pp {\V r}{\V \Delta_2} 
-{\xi_3-\xi_4\over 4}\pp {\V r}({\V \Delta_3}-{\V \Delta_4}) 
\nonumber\\&
={\cal I}_{\rm space}^\xi+{\cal I}_{\rm space}^1
\label{s3}
\end{align}
where we added the first two equations and divided by two to obtain the third equation. The three terms collected in ${\cal I}_{\rm space}^\xi$ will contribute obviously to (\ref{s1}).

As remaining parts in (\ref{balance}) with (\ref{inout}) we consider now the momentum derivatives and $\Delta_K$ and have
\ba
&-{\V \Delta_K\over 2} \left (\pp {\V  k} +\pp {\V p}\right )[D (\xi_3+\xi_4)]
-{D\over 4}{\partial \V \Delta_K\over \partial \omega}
\nonumber\\&
\times\left [
(\xi_1+\xi_2)
\left (\pp {\V k} +\pp {\V p}\right ) E'
+(\xi_3+\xi_4)\left (\pp {\V k} +\pp {\V p}\right ) E\right ].
\label{entk1}
\end{align}
Again we have added the A-transformed expression and divided by 2. 
Replacing further
\ba
{\partial\V \Delta_K\over \partial \omega} \!\! \left (\!\pp {\V k} \!+\!\pp {\V p}\!\right ) \!E\!=\!\left [\!\left (\pp {\V
  k}\!+\!\pp {\V p}\!\right )^{on}\!\!\!-\!\left (\!\pp {\V k}\!+\!\pp {\V p}\!\right )\!\right ]\!\V \Delta_K(E)
\label{onk}
\end{align} 
by the on-shell derivative one gets
\ba
&-{\V \Delta_K\over 2} \left (\pp {\V  k} +\pp {\V p}\right )^{on}[D (\xi_3+\xi_4)]
\nonumber\\&
-{D\over 4}
  (\xi_1+\xi_2)
  \!\left [\!\left (\pp {\V
  k}\!+\!\pp {\V p}\!\right )^{on}\!-\!\left (\pp {\V k}\!+\!\pp {\V p}\right )\right ]\!\V \Delta_K(E')
  \nonumber\\
  &
  -{D\over 4}(\xi_3+\xi_4)\!\left [\!\left (\pp {\V
  k}\!+\!\pp {\V p}\!\right )^{on}\!-\!\left (\pp {\V k}\!+\!\pp {\V p}\right )\right ]\!\V \Delta_K(E).
\label{entk11}
\end{align}
Observing (\ref{relt})
allows to add both last terms in (\ref{entk11}) and we can create the on-shell derivative needed for the first term in (\ref{entk1}) to find
\be
&&-\left (\pp {\V k}+\pp {\V p}\right )^{on} \left [\frac D 2\V \Delta_K (\xi_3+\xi_4)\right ]
\nonumber\\&&
+{\xi_3+\xi_4-\xi_1-\xi_2\over 4}D\left (\pp {\V k}+\pp {\V p}\right )^{on}\V \Delta_K
\nonumber\\&&
+{\xi_1+\xi_2+\xi_3+\xi_4\over 4}D\left (\pp {\V k}+\pp {\V p}\right )\V \Delta_K.
\label{kon}
\ee
The first term vanishes under integration, the second term obviously accounts to the gain and the last term can be rewritten as spatial derivatives of the spatial shifts according to the definition (\ref{vc4g}), i.e. 
\ba
\pp {\V r} \left ({\V \Delta_3}\!+\!{\V \Delta_2}\!-\!{\V \Delta_4}\right )=2 \left (\pp {\V p}\!+\!\pp {\V
  k}\right )\V \Delta_K\!+\!2\pp {\V r} {\V \Delta_3}.
\label{identityR}
\end{align}  
The result for (\ref{kon}) finally reads
\ba
&{\xi_3+\xi_4-\xi_1-\xi_2\over 4}D\left (\pp {\V k}+\pp {\V p}\right )^{on}\V \Delta_K
\nonumber\\&+{\xi_1+\xi_2+\xi_3+\xi_4\over 4}\frac D 2 \pp {\V r} \left
  ({\V \Delta_2}-{\V \Delta_3}-{\V \Delta_4}\right )
\nonumber\\
&={\cal I}_{\rm gain}^K++{\cal I}_{\rm space}^2.
\label{s4}
\end{align}

Now we have all terms in a form to be combined. The spatial derivatives of
$\Delta$s in (\ref{s4}) and (\ref{s3}) can be regrouped together as
\be
&&{\cal I}_{\rm space}^1+{\cal I}_{\rm space}^2={\xi_3+\xi_4-\xi_1-\xi_2\over 8}D \pp {\V r} ({\V \Delta_3}+{\V \Delta_4}+{\V \Delta_2})
\nonumber\\&&
-\frac D 2 \left ({\xi_3}\pp {\V r}{\V \Delta_3} 
+{\xi_4}\pp {\V r}{\V \Delta_4}-{\xi_2}\pp {\V r}{\V \Delta_2}\right )
\nonumber\\&&
={\cal I}_{\rm gain}^\Delta+{\cal I}_{\rm space}^\Delta.
\label{s5}
\ee  
Collecting the terms from (\ref{s1}), (\ref{s2}), (\ref{s3}) and (\ref{s5}) with
spatial gradients which have no $\xi_3+\xi_4-\xi_1-\xi_2$ prefactor,
we obtain the divergence
\be
{\cal I}_{\rm space}^D+{\cal I}_{\rm space}^\omega+{\cal I}_{\rm space}^\xi+{\cal I}_{\rm space}^\Delta=-{\partial\V j_\xi^{\rm mol}\over \partial \V r}
\ee
 of the observable current
\ba
\V j_\xi^{\rm mol}\!=\!-\frac 1 2 \!\sum_{ab}\!\int\!\!{d^3 \! kd^3 \! pd^3 \! q\over(2\pi)^9}
D(\xi_2{\V \Delta_2}\!-\!\xi_3{\V \Delta_3}\!-\!\xi_4{\V \Delta_4})
\label{entcurrmol}
\end{align}
which can now be added to the quasiparticle part from the drift side (\ref{ob51ent}). Obviously it has again the statistical interpretation as the observable per delay time $\xi/\Delta_t$ carried at the points of nonlocal collisions multiplied with the probability to form a molecule $D\Delta_t$.

\subsection{Remaining gains}\label{agains}
The remaining terms with the prefactor $\xi_3+\xi_4-\xi_1-\xi_2$ in
(\ref{ec10ent}), (\ref{s2}), (\ref{s4}) and (\ref{s5}) are of gain form and read 
\ba
&
{\cal I}_{\rm gain}^\xi+{\cal I}_{\rm gain}^\omega+{\cal I}_{\rm gain}^K+{\cal I}_{\rm gain}^\Delta
=\biggl \{
2\left (\!1\!+\!{1\over 2}\pp t\Delta_t\!\right )
\nonumber\\
&\!+\!\left (\pp {\V k}\!+\!\pp {\V p}\right )^{on}\V \Delta_K \!+\!
\frac 1 2 \pp {\V r} ({\V \Delta_3}\!+\!{\V \Delta_4}\!+\!{\V \Delta_2})
\nonumber\\&
\!+\!{\partial \varepsilon_2\over \partial \V r} {\partial {\V \Delta_2}\over \partial \omega}
\!+\!{\partial \varepsilon_3 \over \partial \V r}{\partial {\V \Delta_3}\over \partial \omega}
\!+\!{\partial \varepsilon_4 \over \partial \V r}{\partial {\V \Delta_4}\over \partial \omega}
\!\biggr \}
{\xi_3+\xi_4-\xi_1-\xi_2\over 4}D.
\label{entgain1}
\end{align}
We see that for both observables,
$\xi=k$ for momentum or $\xi=\varepsilon_k$ for energy, the derivative terms are of higher order in
$\Delta$ since the differences in $\xi$s lead to momentum and energy shift
itself, respectively. The zeroth order for momentum and energy gain we had
already shown to combine together with a drift part into derivatives of the
quasiparticle values, (\ref{ec79}) and (\ref{ec11}). The gain for density,
$\xi=1$ vanishes trivially.

Therefore we see that only for the entropy an extra gain term remains from the collision integral. We rewrite the $\{\}$-bracket in (\ref{entgain1}) using (\ref{onk}) and
(\ref{identityR}) to get
\be
\frac 1 2 \{\}&=&1\!+\!{1\over 2}\pp t\Delta_t\!+\!\frac 1 2 {\partial\V \Delta_K\over \partial \omega}  
\left (\pp {\V k} \!+\!\pp {\V p} \right ) E'
\!+\!\frac 1 2 \pp {\V r} \V \Delta_2
\nonumber\\&&
\!+\!\frac 1 2{\partial \varepsilon_2 \over \partial \V r} {\partial {\V \Delta_2}\over \partial \omega}
+\frac 1 2{\partial \varepsilon_3 \over \partial \V r}{\partial {\V \Delta_3}\over \partial \omega}
+\frac 1 2{\partial \varepsilon_4 \over \partial \V r}{\partial {\V \Delta_4}\over \partial \omega}
\label{entgain2}
\ee
where the part $\p {\V r} {\V \Delta_2}$ vanishes when we add the A-transformed
expression and divide by 2.

Now we remember that the weight of the energy conserving $\delta$-function has been used for the symmetrized energies (\ref{ob28}). We had for out-scattering, which is the one contained in $D$, 
\ba
&\left .
\delta
\bigl ( 
\varepsilon_1+\varepsilon_2-\varepsilon_3-\varepsilon_4+2\Delta_E 
\bigr )
\right |_{\omega=\varepsilon_1+\varepsilon_2}
\nonumber\\
&=\! \left [
1\!+\!
{1\over 2}
\left(
{\partial \varepsilon_3\over\partial \V k}
\!+\!{\partial \varepsilon_4\over\partial \V p}
\right)
{\partial{\V \Delta_K}\over\partial\omega}
\!+\!{1\over 2}
\left( \!
-{\partial\varepsilon_2\over\partial \V r}
{\partial{\V \Delta_2}\over\partial\omega}+
{\partial\varepsilon_3\over\partial \V r}
{\partial{\V \Delta_3}\over\partial\omega}
\right . \right .\nonumber\\& \left . \left .
\qquad +
{\partial\varepsilon_4\over\partial \V r}
{\partial{\V \Delta_4}\over\partial\omega}
\!\right)
\!-\!{\partial \Delta_E\over \partial \omega} \!+\!\frac 1 2 {\partial \Delta_t \over \partial \omega}  {\partial E'\over \partial t} 
\right ]
\nonumber\\&
\left .
\qquad \times\delta
\left(
\varepsilon_1\!+\!\varepsilon_2\!-\!\varepsilon_3\!-\!\varepsilon_4
+2\Delta_E\right )
\right |_{\omega=E}
\label{ob28s}
\end{align}
and comparing this with (\ref{entgain2}) we find for the entropy gain (\ref{entgain1})
\ba
\left (1\!+\!{\partial\varepsilon_2\over\partial \V r}
{\partial{\V \Delta_2}\over\partial\omega}\right ){\xi_3\!+\!\xi_4\!-\!\xi_1\!-\!\xi_2\over 2}D_{\varepsilon_1+\varepsilon_2}
\end{align}
where we used (\ref{onshell}).
If we understand the $\delta$-function as selfconsistent solution with respect to the shifted argument of $\epsilon_2$ we can absorb the factor
\ba
\delta \{\omega\!-\!\varepsilon_1\!-\!\varepsilon_2[\V r\!+\!{\V \Delta_2}(\omega)]\}=\delta(\omega\!-\!\varepsilon_1\!-\!\varepsilon_2)
\left (\!1\!+\!{\partial\varepsilon_2\over\partial \V r}
{\partial{\V \Delta_2}\over\partial\omega}\!\right )
\end{align}
and write finally for the entropy gain
\ba
{\cal I}_{\rm gain}^S=\frac 1 2 \sum_{ab}\!\int\!\!{d^3 \! kd^3 \! pd^3 \! q\over(2\pi)^9}
\, {(\xi_3\!+\!\xi_4\!-\!\xi_1\!-\!\xi_2 )}D_{\varepsilon_1}.
\label{gainentropy}
\end{align}
Comparing to (\ref{ec10ent}) the energy-conserving $\delta$-function is now to be understood as selfconsistent expression of shifts. This is required in order to have the same shifts in $\varepsilon$ inside the $\delta$-function as inside the distributions. 

\section{Summary on balance equations and proof of H-theorem}

\subsection{Equation of continuity}

We have found that the density balance equation from the nonlocal kinetic theory 
consists of quasiparticle parts and molecular contributions
\begin{equation}
{\partial (n^{\rm qp}_a+n^{\rm mol}_a)\over\partial t}+{\partial (\V {j}^{\rm qp}_a+\V j^{\rm mol}_a)\over\partial \V r}=0
\label{ob1}
\end{equation}
with the standard quasiparticle density (\ref{ob5}) and current (\ref{ob51a}).
The correlated or molecular density (\ref{phicurr})
\ba
n_a^{\rm mol}&=\sum_b\int{d^3 \! kd^3 \! pd^3 \! q\over(2\pi)^9}
|t_{\rm sc}(\varepsilon_1+\varepsilon_2,k,p,q)|^2\Delta_t
\nonumber\\
&\times 2\pi\delta(\varepsilon_1+\varepsilon_2-\varepsilon_3-
\varepsilon_4)f_3f_4(1-f_1-f_2)
\label{ob25}
\end{align}
has the statistical interpretation of the rate of binary processes $D$ of (\ref{ec63}) weighed with the $\Delta_t$. 

The molecular current (\ref{entcurrmol}) we have obtained as
\ba
\V j_a^{\rm mol}\!=\!\sum_b\!\int\!\!{d^3 \! kd^3 \! pd^3 \! q\over(2\pi)^9}{\V \Delta_3}
D
.
\label{ob60}
\end{align}
Applying transform A, add and dividing by two we can write equivalently in (\ref{ob60}) for $\V \Delta_3$ also  $\V \Delta_{\rm fl}$ of (\ref{combi}).
Again we obtain a statistical interpretation in that the velocity of the molecule ${\V \Delta_{\rm fl}}/\Delta_t$ is multiplied with the rate $D$ to form a molecule and weighted with the duration $\Delta_t$.

\subsection{Energy balance}

The energy balance (\ref{balanceEa}) we found as
\be
{\partial ({\cal E}^{\rm qp}+{\cal E}^{\rm mol})\over \partial t}+{\partial (\V j_E^{\rm qp}+\V j_E^{\rm mol})\over \partial \V r}=0
\label{balanceEb}
\ee
with the quasiparticle energy functional 
(\ref{ec22cb})
having the same structure as the uncorrelated energy functional,
the 
bare interaction potential is, however, replaced by the T-matrix.

The molecular contribution to the energy (\ref{phicurr}) 
\ba
{\cal E}^{\rm mol}={1\over 2}\sum_{ab}\int{d^3 \! kd^3 \! pd^3 \! q\over(2\pi)^9}
D
\Delta_t E,
\label{ec8aa}
\end{align}
has also a natural statistical interpretation. The factor $D \Delta_t$ measures the probability of finding two particles in the scattering state. The total
energy of these two particles is the mean of $E=\varepsilon_1+\varepsilon_2$.

The energy current is the sum of the quasiparticle current (\ref{ob51e}) 
and the molecular current (\ref{entcurrmol})
\ba
\V j_E^{\rm mol}=\frac 1 2 \sum_{ab}\!\int\!\!{d^3 \! kd^3 \! pd^3 \! q\over(2\pi)^9}D(\varepsilon_2 {\V \Delta_2}\!-\!\varepsilon_3 {\V \Delta_3}\!-\!\varepsilon_4 {\V \Delta_4}).
\label{ecurrmol}
\end{align}
It is the balance of energies carried by the different spatial off-sets.

\subsection{Navier-Stokes equation}
The inertial force density is given by the time derivative of the
momentum density ${\cal Q}$. The deformation force density is
given by the divergence of the stress tensor.
The stress tensor we derived from the balance between the inertial
and the deformations forces
\begin{equation}
{\partial \left ({\cal Q}_j^{\rm qp}+{\cal Q}_j^{\rm mol}\right )\over\partial t}=-\sum_i{\partial \left ( {\Pi}^{\rm qp}_{ij}+{\Pi}^{\rm mol}_{ij}\right )\over
\partial r_i}
\label{ec51aa}
\end{equation}
with the momentum density consisting of the quasiparticle (\ref{ec51ab}) and molecular part (\ref{phicurr}) with $\xi=k_j$
\ba
&{\cal Q}_j^{\rm mol}
={1\over 2}\!\sum_{ab}\!\int{d^3 \! kd^3 \! pd^3 \! q\over(2\pi)^9}\!
(k_j\!+\!p_j)D
\Delta_t
\label{ec55ac}
\end{align}
which gives the mean momentum carried by a molecule formed with the rate $D$ and lifetime $\Delta_t$.

Observing (\ref{ec79}), the total stress tensor formed by the quasiparticles read
\begin{equation}
{\Pi}_{ij}^{\rm qp}=\sum_a\int{d^3 \! k\over(2\pi)^3}\left(k_j
{\partial\varepsilon\over\partial k_i}+\delta_{ij}\varepsilon\right)f-
\delta_{ij}{\cal E}^{\rm qp}
\label{ec81}
\end{equation}
with (\ref{ec22cb}) and the
collision-flux contributions (\ref{entcurrmol}),
\ba
{\Pi}_{ij}^{\rm mol}=&{1\over 2}\sum_{ab}\int{d^3 \! kd^3 \! pd^3 \! q\over(2\pi)^9}
D
\nonumber\\&\times
\left [(k_j-q_j) \Delta_{3i}+(p_j+q_j )\Delta_{4i}-p_j \Delta_{2i}\right ]
\label{ec82}
\end{align}
is the correlated part of the stress tensor.
It possesses a statistical interpretation as well. The two-particle state is characterized
by the initial momenta $\V k$ and $\V p$ and the transferred momentum $\V q$. The momentum tensor is the balance of the momenta carried by the corresponding spatial off-sets weighted with the rate to form a molecule $D$.

Let us comment here on the novelty of the results. The correlated density (\ref{ob25}) and molecular current (\ref{ob60}) as well as the molecular contribution to the energy (\ref{ec8aa}) and stress tensor (\ref{ec82}) have been first derived in \cite{LSM97}. The molecular energy current (\ref{ecurrmol}) as well as the following entropy balance and the H-theorem are new results obtained here with the help of the transformations A and B.
 
\subsection{Entropy balance}
Finally the entropy balance reads
\be
{\partial ({\cal S}^{\rm qp}+{\cal S}^{\rm mol})\over \partial t}+{\partial (\V j_S^{\rm qp}+\V j_S^{\rm mol})\over \partial \V r}={\cal I}_{\rm gain}^S.
\label{balanceEc}
\ee
where the entropy consists of the quasiparticle part 
(\ref{entropquasi})
and the molecular part (\ref{phicurr})
\ba
{\cal S}^{\rm mol}=&-{k_B\over 2}\sum_{ab}\int{d^3 \! kd^3 \! pd^3 \! q\over(2\pi)^9}
|t_{\rm sc}|^2\Delta_t 2\pi\delta(\varepsilon_1\!+\!\varepsilon_2\!-\!\varepsilon_3\!-\! \varepsilon_4)
\nonumber\\
&\times f_1f_2(1-f_3-f_4)\ln{f_3 f_4\over  (1-f_3)(1-f_4) }.
\label{entcurr}
\end{align}
In the same way, the entropy current has a quasiparticle part (\ref{ob51ent})
and a molecular contribution (\ref{entcurrmol}) with $\xi_1=-k_B\ln f_1/(1-f_1)$
reading
\ba
&\V j_S^{\rm mol}=
\frac{k_B}{2} \sum_{ab}\int{d^3k d^3p d^3q\over(2\pi)^9} D
\nonumber\\
&\times \!\left [\ln {f_2\over (1\!-\!f_2)}{\V \Delta_2}\!-\!\ln {f_3\over (1\!-\!f_3)}{\V \Delta_3}
  \!-\!\ln {f_4\over (1\!-\!f_4)}{\V \Delta_4}\right ].
\label{entcurrmolS}
\end{align}
The entropy gain (\ref{gainentropy}) finally reads
\ba
&{\cal I}_{\rm gain}^S
=
-\frac{k_B}{2} \sum_{ab}\int{d^3 \! kd^3 \! pd^3 \! q\over(2\pi)^9}
f_1f_2(1\!-\!f_3-\!f_4)
\nonumber\\
&\times 
2\pi\delta(\varepsilon_1\!+\!\varepsilon_2\!-\!\varepsilon_3\!-\! \varepsilon_4)|t_{\rm sc}|^2
\ln {{f_3 f_4 (1\!-\!f_1)(1\!-\!f_2)\over (1\!-\!f_3)(1\!-\!f_4) f_1 f_2}}.
\label{ec10enta}
\end{align}
This entropy gain remains explicit while the momentum gain and energy gain are transferring kinetic into correlation parts and do not appear explicitly.

\subsection{Proof of H-theorem}

Now we are going to proof that the entropy gain 
(\ref{gainentropy}) or (\ref{ec10enta}) is always positive. We consider in short-hand notation
$\xi =\xi_3+\xi_4-\xi_1-\xi_2$. Then the expansion in $\Delta$s reads
\ba
&{\cal I}_{\rm gain}^S=\frac D 2 \xi=\biggl [
1  +\Delta_t \left ({\partial^3\over \partial t}+{\partial^4\over \partial t}\right )
+\V \Delta_2{\partial^2\over \partial{\V r}}+\V \Delta_3{\partial^3\over \partial{\V r}}
\nonumber\\
&
+\V \Delta_4{\partial^4\over \partial{\V r}}
+\V \Delta_K \left ({\partial^3\over \partial {\V k}}+{\partial^3\over \partial {\V p}} +{\partial^4\over \partial{\V k}}+{\partial^4\over \partial {\V p}}\right )
\biggr ]
{D_0\over 2}\xi_0
\end{align}
where we indicate explicitly to which argument $1,2,3$ or $4$ the derivatives apply.

First we establish a useful relation and focus on the time derivatives. Let us consider an unknown derivative operator ${\cal R}$ and apply transform B together with the space-and time reversal transformation inverting the shifts
\ba
&(1+{\cal R})\frac D 2 \xi=
\left [
1+{\cal R}+\Delta_t 
\left ({\partial^3 \over \partial t}+{\partial^4 \over \partial t}\right )
\right ]
{D_0\over 2}\xi_0
\nonumber\\
&=\frac 1 2 \left [
-1-\tilde {\cal R}_B-\Delta_t 
\left ({\partial^1 \over \partial t}+{\partial^2\over \partial t}\right )
\right ]
(I_0+D_0)\xi_0
\end{align}
where we denote the symmetrized collision term $I_0=2\pi \delta (E-E') |t_{\rm sc}|^2[f_3f_4(1-f_1-f_2)-f_1f_2(1-f_3-f_4)]$. Subtracting the $D_0$ part from the left we obtain
\ba
&\left [
1+{{\cal R}+\tilde {\cal R_B}\over 2 }+{\Delta_t \over 2}
\left ({\partial^1 \over \partial t}+{\partial^2 \over \partial t}-{\partial^3 \over \partial t}-{\partial^4 \over \partial t}\right )
\right ]
\frac D 2 \xi
\nonumber\\
&=\frac 1 4 \left [
-1-\tilde {\cal R}_B-\Delta_t \left ({\partial^1 \over \partial t}+{\partial^2 \over \partial t}\right )\right ]I_0 \xi_0.
\label{rrb}
\end{align}
Now we determine the unknown derivative operator ${\cal R}=\Delta_t(a\partial^2+b\partial^2+c\partial³+d\partial^4)$ and consequently $\tilde {\cal R}_B=-\Delta_t(c\partial^2+d\partial^2+a\partial³+b\partial^4)$ such that the left hand side of (\ref{rrb}) is unity, which provides $a=b=-1$ and $c=d=0$ and we obtain finally the identity
\be
{\cal I}_{\rm gain}^S=\frac D 2 \xi=-\frac 1 4 \left (1+\Delta_t{ \partial \over \partial t}\right ) I_0 \xi_0
\label{theo1}
\ee
suited for proving the H-theorem. Replacing the time derivative by the momentum derivative and $\Delta_t$ by $\Delta_K$ we obtain the analogous expression. It is worth to show how the ${\cal R}$ operator looks like for these spatial derivatives. Analogously to (\ref{rrb}) we have
\ba
&\left [\!
1\!+\!{{\cal R}\!+\!\tilde {\cal R_B}\over 2 }\!-\!\V\Delta_2 {\partial^2 \over \partial {\V r}}\!+\!{\V \Delta_3\over 2} \left ({\partial^1 \over \partial {\V r}}\!+\!{\partial^2 \over \partial {\V r}}\!-\!{\partial^3 \over \partial {\V r}}\!+\!{\partial^4 \over \partial {\V r}}\right )
\right .
\nonumber\\
&\qquad\left . \!-\!\V \Delta_4 {\partial^4 \over \partial {\V r}}
\right ] \frac D 2 \xi
\nonumber\\
=&-\left [
1\!+\!\tilde {\cal R}_B
\!-\!\V\Delta_2 {\partial^2 \over \partial {\V r}}\!+\!{\V \Delta_3} \left ({\partial^1 \over \partial {\V r}}+{\partial^2\over \partial {\V r}}+{\partial^4 \over \partial {\V r}}\right )\!-\!\V \Delta_4 {\partial^4 \over \partial {\V r}}
\right ]
\nonumber\\
&\qquad\times{I_0\over 4} \xi_0.
\label{rrbx}
\end{align}
Again we search for an operator ${\cal R}$ which renders the left side unity. A linear equation system provides a manifold of solutions from which we choose one
with the final result together with (\ref{theo1})
\ba
{\cal I}_{\rm gain}^S=&\frac D 2 \xi=-\left [
1\!+\!\Delta_t{\partial \over \partial t}\!-\!\V\Delta_2 {\partial^2 \over \partial {\V r}}\!-\!{\V \Delta_3} {\partial^3 \over \partial {\V r}}\!-\!\V \Delta_4 {\partial^4 \over \partial {\V r}}
\right .
\nonumber\\
&\qquad\left .\!+\!\V \Delta_K\left ( {\partial \over \partial {\V k}}+{\partial \over \partial {\V p}}\right )\right ] {I_0\over 4} \xi_0
\nonumber\\
=&{k_B \over 4}  \sum_{ab}\!\!\int{d^3 \! \!kd^3 \! pd^3 \! q\over(2\pi)^9}
2\pi\delta(\varepsilon_1\!+\!\varepsilon_2\!-\!\varepsilon_3\!-\! \varepsilon_4)|t_{\rm sc}|^2
\nonumber\\
&\times
\biggl \{
f_3f_4(1\!-\!f_1)(1\!-\!f_2)-
f_1f_2(1\!-\!f_3)(1\!-\!f_4)
\biggr \}
\nonumber\\
&\times
\left .\ln {{f_3 f_4 (1\!-\!f_1)(1\!-\!f_2)\over (1\!-\!f_3)(1\!-\!f_4) f_1 f_2}}\right |_{1,2,3,4\, {\rm equally}\,{\rm shifted}}
\end{align}
where we reestablished the full notation.
This entropy gain is always positive since with $a=f_3f_4(1-f_1)(1-f_2)$ and $b=f_1f_2(1-f_3)(1-f_4)$ we have the always positive entropy production density $(a-b)\ln(a/b)>0$. This is completely analogously to the proof of Boltzmann's H-theorem.  


We therefore have shown that the second law of thermodynamics holds also within the nonlocal kinetic theory. We want to emphasize that the molecular contribution to the entropy due to particle interactions as well as the correlated entropy current are new results and show how the two-particle correlations exceed the Landau theory. The single-particle entropy can decrease on cost of the molecular part of entropy describing the two-particles in a molecular state.

In a forthcoming paper one should show how the here obtained results compare to the results from various microscopic approaches \cite{CP75,VB98,MT15,BIR01}. Since this requires extensive algebra to reduce the general expressions in the literature to transparent forms in terms of distribution functions and phase shifts as presented here, we feel that this exceeds the possibility of one paper. Here we restrict therefore to show the consequences of the nonlocal kinetic equation to quantum hydrodynamical equations providing a thermodynamically consistent set.

\section{Conclusion}

We have presented the molecular parts of observables and their currents which add to the known quasiparticle expressions. These contributions emerge from the nonlocal kinetic equation.
The molecular parts to the observables possess a statistical interpretation that the rate to form a two-particle molecule is multiplied with the lifetime of the molecule and with the observable. The currents are correspondingly the observables per lifetime carried through the nonlocal collision which possess the virial form.

All nonlocal shifts are possible to calculate as derivatives of the phase shift of the scattering T-matrix. While the modulus of the T-matrix determines the cross section, the phase provides the nonlocal picture. Due to the correlated or molecular parts of the observables the nonlocal scenario exceeds the Landau quasiparticle theory to which it collapses in local approximation. Since the nonlocal picture leads immediately to the virial correlations, the Enskog extensions of Boltzmann equation for dense gases are combined with the Landau theory of quasiparticles in this nonlocal kinetic equation.

The thermodynamical quantities are expressed in nonequilibrium form. The necessary distributions as solutions of the kinetic equation can be time-dependent as well as the shifts arising from the T-matrix. The latter ones as solution of Bethe-Salpeter equation are time dependent themselves. This time-dependence leads to an energy gain which is the correlation energy transferred from or to the system during the collision. We have used
here that this energy gain combines with the quasiparticle kinetic energy rate into a rate of total quasiparticle energy as it was proved earlier \cite{LSM97}. In this sense there is a continuous transfer of correlation energy to kinetic energy preserving the total energy. The same transfer happens for the mean momentum as force supplied continuously from or to the system during the collision leading to the momentum tensor in agreement with the total energy of the system.

This consistent picture is completed by the entropy. The entropy balance reveals a gain term which is proved to be larger zero such that the H-theorem holds also for the nonlocal quantum kinetic theory and therefore the second law of thermodynamics. We found the explicit expression for the molecular contribution to the decrease of local entropy production if a molecule is formed with a certain lifetime. The molecular entropy adds to the known Landau form of quasiparticles and is supposed to have many applications. In the theory of cold gases one might think of the contribution of short living bound states which can be described in this way. In nuclear physics the short-living resonances are described herewith within a consistent transport theory. Finally the correlation contribution to the viscosity will become feasible \cite{ACHMN08}.

\acknowledgments
P. Lipavsk\'y is gratefully thanked for helpful critical comments. Many discussions with {\v V}. {\^ S}pi{\^c}ka in the early history of this nonlocal kinetic theory is acknowledged.

\bibliography{bose,kmsr,kmsr1,kmsr2,kmsr3,kmsr4,kmsr5,kmsr6,kmsr7,delay2,spin,spin1,refer,delay3,gdr,chaos,sem3,sem1,sem2,short,cauchy,genn,paradox,deform,shuttling,blase,spinhall,spincurrent,tdgl,pattern,zitter,graphene}
\bibliographystyle{aipnum4-1}
\end{document}